\documentclass[apj]{emulateapj}
\usepackage{color}
\usepackage[usenames,dvipsnames]{xcolor}
\usepackage{epsfig}
\usepackage{graphicx}
\usepackage{amsmath}
\usepackage{amsfonts}
\usepackage{amssymb}
\usepackage{colordvi}
\usepackage{float}

\newcommand{\Mg}{Mg$_{UV}$ }
\newcommand{\OII}{{\rm [OII]}}
\newcommand{\lya}{{\rm Ly$\alpha$ }}
\newcommand{\hal}{{\rm H$\alpha$ }}
\def\EW{\ifmmode{\mathrm{EW}_\mathrm{rf}\mathrm{([OII])}}\else{EW$_\mathrm{rf}$([OII])}\fi}

\begin{document}

\shorttitle{SHARDS ELGs at {\it z}\,$\sim$0.84 and {\it z}\,$\sim$1.23}
\shortauthors{A. Cava et al.}
      
\slugcomment{Submitted to ApJ}

\title{SHARDS: a global view of the star formation activity at \MakeLowercase{{\it z}}\,$\sim$0.84 and \MakeLowercase{{\it z}}\,$\sim$1.23}

\author{Antonio~Cava\altaffilmark{1}, Pablo~G.~P{\'e}rez-Gonz{\'a}lez\altaffilmark{2}, M.~Carmen~Eliche-Moral\altaffilmark{2}, Elena~Ricciardelli\altaffilmark{3}, Alba~Vidal-Garc\'ia\altaffilmark{4}, Belen~Alcalde Pampliega\altaffilmark{2}, Almudena~Alonso-Herrero\altaffilmark{5}, Guillermo~Barro\altaffilmark{6}, Nicolas~Cardiel\altaffilmark{2}, A.~Javier~Cenarro\altaffilmark{7}, Stephane~Charlot\altaffilmark{4}, Emanuele~Daddi\altaffilmark{8}, Miroslava~Dessauges-Zavadsky\altaffilmark{1}, Helena~Dom\'inguez~S\'anchez\altaffilmark{2}, Nestor~Espino-Briones\altaffilmark{2}, Pilar~Esquej\altaffilmark{2}, Jesus~Gallego\altaffilmark{2}, Antonio~Hern\'an-Caballero\altaffilmark{5}, Marc~Huertas-Company\altaffilmark{9}, Anton~M.~Koekemoer\altaffilmark{10}, Casiana~Mu\~noz-Tunon\altaffilmark{11}, Jose~M.~Rodriguez-Espinosa\altaffilmark{11}, Lucia~Rodr\'iguez-Mu\~noz\altaffilmark{2,12}, Laurence~Tresse\altaffilmark{13},  Victor~Villar\altaffilmark{14}}

\affil{$^1$ Observatoire de Gen{\`e}ve, Universit{\'e} de Gen{\`e}ve, 51 Ch. des Maillettes, 1290 Versoix, Switzerland}
\affil{$^2$ Departamento de Astrof\'{i}sica y Ciencias de la Atm\'osfera, Facultad de CC. F\'{\i}sicas, Universidad Complutense de Madrid, E-28040, Madrid, Spain}
\affil{$^3$ Departament d'Astronomia i Astrofisica, Universitat de Valencia, c/ Dr. Moliner 50, E-46100 - Burjassot, Valencia, Spain}
\affil{$^4$ Institut d'Astrophysique de Paris, CNRS, Universit\'e Pierre \& Marie Curie, UMR 7095, 98bis bd Arago, 75014 Paris, France}
\affil{$^5$ Instituto de F\'isica de Cantabria, CSIC-UC, Avenida de los Castros s/n, E-39005 Santander, Spain}
\affil{$^6$ UCO/Lick Observatory, Department of Astronomy and Astrophysics, University of California, Santa Cruz, CA 95064, USA}
\affil{$^7$ Centro de Estudios de F\'isica del Cosmos de Arag\'on (CEFCA), Plaza San Juan 1, 44001 Teruel, Spain}
\affil{$^8$ CEA, Laboratoire AIM, Irfu/SAp, F-91191 Gif-sur-Yvette, France}
\affil{$^9$ University Denis Diderot, CNRS, GEPI-Observatoire de Paris UMR 8111, Paris, France}
\affil{$^{10}$ Space Telescope Science Institute, 3700 San Martin Dr., Baltimore, MD 21218}
\affil{$^{11}$ Instituto de Astrof\'{\i}sica de Canarias, 38200 La Laguna, Tenerife; Departamento de Astrof\'{\i}sica, Universidad de   La Laguna, E-38205 La Laguna, Tenerife, Spain}
\affil{$^{12}$ Dipartimento di Fisica e Astronomia "G. Galilei", Universit\'a di Padova, Vicolo dell'Osservatorio 3, I-35122, Italy}
\affil{$^{13}$ Aix Marseille Universit\'e, CNRS, LAM (Laboratoire d'Astrophysique de Marseille) UMR 7326, 13388, Marseille, France}
\affil{$^{14}$ European Space Astronomy Centre, PO Box 78, E-28691 Villanueva de la Canada, Madrid, Spain}

\label{firstpage}
\begin{abstract}
In this paper, we present a comprehensive analysis of star-forming galaxies (SFGs) at intermediate redshifts ({\it z}\,$\sim$1). We combine the ultra-deep optical spectro-photometric data from the SHARDS survey with deep UV-to-FIR observations in the GOODS-N field. Exploiting two of the 25 SHARDS medium-band filters, F687W17 and F823W17, we select \OII\ emission line galaxies (ELGs) at {\it z}\,$\sim$0.84 and {\it z}\,$\sim$1.23 and characterize their physical properties.
Their rest-frame equivalent widths (\EW), line fluxes, luminosities, star formation rates (SFRs) and dust attenuation properties are investigated. The evolution of the \EW\ closely follows the SFR density evolution of the Universe, with a \EW$\propto$(1+{\it z})$^3$ trend up to redshift {\it z}\,$\simeq$1, followed by a possible flattening. 
The SF properties of the galaxies selected on the basis of their [OII] emission are compared with complementary samples of SFGs selected  by their MIR and FIR emission, and also with a general mass-selected sample of galaxies at the same redshifts. 
We observationally demonstrate that the UVJ diagram (or, similarly, a cut in the specific SFR) is only partially able to distinguish the quiescent galaxies from the SFGs.  The SFR-M$_*$ relation is investigated for the different samples, finding a logarithmic slope $\sim$1, in good agreement with previous results.
The dust attenuations derived from different SFR indicators (UV(1600), UV(2800), [OII], IR) are compared, finding clear trends with respect to both the stellar mass and total SFR, with more massive and highly star-forming galaxies being affected by stronger dust attenuation.
\end{abstract}

\section{Introduction}
\label{sect:intro}

\setcounter{footnote}{0}

The study of emission line galaxies (ELGs) has long been recognized as one of the most powerful and direct tools to investigate star-forming galaxies (SFGs) at different redshifts.  For local SFGs, the most common and reliable emission line used to investigate star formation (SF) is H$\alpha$ \cite[see e.g.][]{1995ApJ...455L...1G,1998ApJ...495..691T,2003ApJ...591..827P,2012MNRAS.426..330D}. 
At increasing redshifts other lines and methods must be used because the H$\alpha$ line runs out of the optical and near infrared (NIR) range, becoming more difficult to be observed. Present day NIR facilities allow to extend these local studies to higher redshifts using spectroscopy \citep[see e.g.][]{2002MNRAS.337..369T,2012MNRAS.420.1061T} or narrow-band photometry \citep[see, e.g.][]{2008ApJ...677..169V,2011ApJ...740...47V,2009MNRAS.398...75S,2011MNRAS.411..675S,2013MNRAS.428.1128S,2011ApJ...726..109L}, although it remains hard and expensive, in terms of observing time, to obtain deep and wide surveys of line emitters with these instruments.

One of the most popular SF indicators used at intermediate/high redshift is the \OII\ emission line at 3727\,\AA. This line lies at optical wavelengths up to {\it z}\,$\sim$1.5, so it has been widely use to detect and study SFGs for decades. The physical mechanism responsible for the production of this intense emission line (actually a doublet at 3726\,\AA~and 3729\,\AA) has been investigated and well understood for a long time \citep{1927MNRAS..88..134E}, but the relation between environmental conditions (density, temperature, metallicity, ionization state of the gas, ...) and the intrinsic emitted flux is still far from being completely characterized. Nonetheless, \OII\ has demonstrated to be a very useful SFR indicator and empirical calibrations are commonly used to transform \OII\ luminosities into SFR \citep{1989AJ.....97..700G,1998ARA&A..36..189K,1998ApJ...504..622H,2001ApJ...551..825J,2004AJ....127.2002K,2006ApJ...648..281Y,2006ApJS..164...81M,2010MNRAS.409..421G,2010MNRAS.405.2594G,2011MNRAS.414..304G,2013MNRAS.430.1042H}.

The Survey for High-z Absorption Red and Dead Sources (SHARDS\footnote{The SHARDS web page is available at: http://guaix.fis.ucm.es/$\sim$pgperez/SHARDS/}, \citealt{2013ApJ...762...46P}) is an ESO/GTC Large Program carried out with the OSIRIS instrument on the 10.4m Gran Telescopio Canarias (GTC). Our survey has obtained imaging data in the GOODS-N field through 25 medium-band filters covering the wavelength range between 500 and 950~nm in sub-arcsec seeing conditions. The typical width of our filters is 15~nm for wavelengths bluer than 880~nm, and 25--35~nm for the three reddest filters. In each filter, SHARDS is able to detect 26.5 mag sources at the 3$\sigma$ level (at least). SHARDS allows us to study any isolated emission line falling within its wavelength coverage. More specifically, for the \OII\ line, we can cover the interval from {\it z}\,$\sim$0.3 up to {\it z}\,$\sim$1.6. 
In this work, we focus on the analysis of ELGs selected using two of the twenty five SHARDS filters, the F687W17 and F823W17, corresponding to ELGs at z$\simeq$0.84 and z$\simeq$1.23 respectively. We have chosen these two filters because they are representative enough of the whole dataset and they provide two samples of galaxies that are sufficiently separated in redshift ($\sim$1.5~Gyr in time) to allow an investigation of the possible evolutionary effects on the derived physical properties. A more extensive analysis including all the SHARDS filters will be presented in a future work.

One of the aims of this work is to demonstrate the the power of the SHARDS medium-band data to select ELGs down to very faint continuum magnitudes and line fluxes (similar or even fainter that those characteristic of the deepest spectroscopic and narrow-band surveys) and study their physical properties in detail. Even though the equivalent spectral resolution of SHARDS (R$\sim$50) is smaller than the one reached by typical spectroscopic (e.g. TKRS, VVDS, DEEP3, HETDEX or MOSDEF surveys; \citealt{2004AJ....127.3121W,2005A&A...439..845L,2011ApJS..193...14C,2011ApJS..192....5A,2015ApJS..218...15K}) or narrow band surveys (\citealt{2008ApJ...677..169V,2011ApJ...740...47V,2009MNRAS.398...75S,2010MNRAS.404.1551S,2011MNRAS.411..675S,2012MNRAS.420.1926S,2013MNRAS.428.1128S}), the depth and quality of the SHARDS data compensate this a priori disadvantage allowing us to study stellar mass complete samples down to M$_*\sim$10$^9$~M$_{\mathrm{sun}}$ at {\it z}\,$\sim$1 \citep{2014MNRAS.443.3538H}.

In fact, our medium-band survey can reach deeper magnitudes (down to R$\sim$26-27) with respect to typical spectroscopic survey (limited to R$\sim$24-25) by investing substantially lower amount of observing time and overcoming the issues arising when dealing with slit apertures and the limited multiplexing of spectrographs. On the other hand, integral field unit (IFU) spectrographs can also observe all the sources within their field of view (FOV), but this is normally very limited, while the SHARDS filters apply to the relatively wide OSIRIS@GTC field of view (i.e. $\sim7'\times8'$). Furthermore, ultra-deep medium-band selected SF galaxy samples should be less affected by Malmquist bias with respect to spectroscopically observed samples which tend to favour highly SFGs. We also remark that systematic effects are expected to be introduced in the study of emission-line luminosity functions by the typical selection of targets in spectroscopic surveys, which are typically based on magnitudes cuts (i.e., continuum dominated) and may miss faint galaxies with strong emission lines (i.e., large EWs). These biases can also affect the study of the SFR-M$_*$ relation since at low masses the success rate of getting redshifts may be higher if the SFR is above average whereas it may be lower for massive heavily attenuated galaxies \citep[see, e.g.,][ for a discussion of the biases and selection effects on the SFR-M$_*$ relation]{2014MNRAS.443...19R,2014ApJS..214...15S}, thus introducing a flattening in the SFR-M$_*$ relation.  Medium- and narrow-band surveys are, by nature, less prone to this kind of selection biases.

The so called "main sequence" \citep{2007ApJ...660L..43N} of SF galaxies, has been extensively studied in the past using different sample selections from low- \citep[see, e.g.,][]{2004MNRAS.351.1151B,2007A&A...468...33E,2015A&A...579A...2I} to high-redshift \citep[see, e.g.,][]{2014ApJS..214...15S,2014ApJ...791L..25S,2015A&A...575A..74S,2015ApJ...799..183S}, in different environments, ranging from extremely low-density \citep[known as cosmic voids; see ][]{2014MNRAS.445.4045R} to the most crowded regions of the Universe \citep[i.e. filaments and galaxy clusters; see e.g.,][]{2013MNRAS.434..423K,2014ApJ...796...51D,2010ApJ...710L...1V,2015ApJ...798...52V}, and using various diagnostics as SFR indicators \citep[see, e.g.,][]{2014MNRAS.443...19R,2015ApJ...804..149S,2015arXiv150703017S}. Despite the advancements on this topic, the debate on the nature and characteristics of the galaxies populating the main sequence is still open  \citep[see, e.g.,][for a different interpretation of the star-forming main sequence]{2014arXiv1406.5191K}. As indicated above, the determination of the slope and the scatter of this relation and the comparison between different works is made difficult by various systematic effects (e.g., differences in calibrations, selection biases, used SFR indicator, among others; see, \citealt{2014ApJS..214...15S} for a more complete discussion of these uncertainties). Furthermore, to disentangle between differences in the determination of the main sequence due to the mentioned systematic effects, and pure evolutionary effects can be even more subtle. 

The main aims of this paper are: (1) to exploit the ultra-deep SHARDS data to select and investigate intermediate-redshift ELGs on the basis of their \OII\ emission and (2) to perform an (almost) unbiased study of the SF activity of galaxies at intermediate redshifts ($\sim$1). This goal is attained by combining ultra-deep medium-band and multi-wavelength data (from UV to FIR) and comparing various SF indicators (UV, \OII\, IR) to assess the robustness of the results. In particular we focus on (3) the characterization of  SFR-M$_*$ relation and (4) the dependence of the dust attenuation on the physical galaxy properties (M$_*$ and total SFR).
The depth and quality of SHARDS data allow to improve upon previous intermediate redshift studies going down to fainter magnitudes and limiting the effect of the selection biases typical of spectroscopic surveys.

The scheme of this paper follows. In Section~\ref{sect:select}, we summarise the main characteristics of the survey and the data used
in this work. We introduce the sample selection for \OII\ galaxies and we define the different complementary samples (general stellar mass-selected sample, IR detected and UVJ/quiescent galaxies) used throughout the paper. We present the methods used to select ELGs, their redshift distribution and discuss the adopted procedures and caveats.  In Section~\ref{sect:oii-prop}, we present the basic measurements for the equivalent widths and line fluxes characterizing the \OII\ population and discuss their observational properties.  
In Section~\ref{sect:oii-sfr}, we discuss the \OII-based SFRs (SFR(\OII)), the stellar mass-SFR relation, and their stellar population properties (mass and age distributions). In Section~\ref{sect:comp}, we compare the properties for different sample selection and discuss the derived SFR-M$_*$ relations and the differences induced by each selection.  Dust extinction and its dependence on the stellar mass and the SFR are investigated in Section~\ref{sect:dust}. Main results and conclusions are summarised in Section~\ref{sect:concl}.

Throughout this paper we use AB magnitudes and the Chabrier (2003) initial mass function (IMF). We adopt the cosmology $H_{0}=70$ km~s$^{-1}$Mpc$^{-1}$, $\Omega_{m}=0.3$, and $\Omega_{\lambda}=0.7$.

\section{Data and sample selection}
\label{sect:select}
In this work, we have gathered a complete sample of star-forming galaxies at {\it z}\,$\sim$0.84 and {\it z}\,$\sim$1.23 in the GOODS-N field by combining 3 different selections: a pure stellar mass-selected sample of galaxies with SFRs calculated through the rest-frame UV emission at 150-250~nm, an emission-line selection based on the detection of the \OII$\lambda$3727 line with data from the SHARDS \citep{2013ApJ...762...46P}, and a selection of obscured SFGs based on mid- and far-IR data from {\it  Spitzer} and {\it Herschel}. In the following subsections, we present the details about the different datasets used in the selection of SFGs and their characterization. We start with the description of the most unique selection among the three mentioned before: the \OII\ sample selected with the new SHARDS data (see Sections~\ref{sect:data}--\ref{sect:datasets}). Then, we complement this sample with the other selections and perform a comparison of the properties of the sample of \OII\ emitters with the dust-obscured SFGs selected from IR data and the mass-selected sample (characterized by the UV emission, see Sections~\ref{sect:complem}--\ref{sect:AGN}). This approach gives us a comprehensive view of the typical limitations inherent to the different selection methods and highlights the complementarity of the different selected samples. As we will discuss later, the construction of the complementary samples must take into account the fact that the \OII\ emitters selected with SHARDS have a very specific redshift distribution. Therefore, both the mass-selected and the obscured
SFG samples should follow similar redshift distributions. This is mandatory to be able to make a fair comparison between different samples and understand the SF and dust extincion properties of the whole population of SFGs at {\it z}\,$\sim$0.84 and {\it z}\,$\sim$1.23, which we will discuss in Sections~\ref{sect:oii-sfr}, \ref{sect:comp} and \ref{sect:dust}.

\subsection{Datasets gathered for this work}
\label{sect:data}
\begin{deluxetable*}{lcccccccccc}[tb]
\tabletypesize{\footnotesize}
\tablecaption{\label{table:filters}Characteristics of the SHARDS filters and observations used in this study}
\tablehead{ \colhead{Filter} & \colhead{CWL} & \colhead{Width}  & \multicolumn{2}{c}{Exposure time}& & \multicolumn{2}{c}{m$_\mathrm{3\sigma}$} & & \multicolumn{2}{c}{seeing}\\
\cline{4-5}                          \cline{7-8}                                \cline{10-11}        \\
&  & &  P1 &  P2 & & P1 & P2 & & P1 & P2 \\
\colhead{(1)} & \colhead{(2)} & \colhead{(3)} &  \colhead{(4)} &  \colhead{(4)} & & \colhead{(5)} & \colhead{(5)} & & \colhead{(6)} & \colhead{(6)}}
\startdata
F670W17  & 670.4 & 16.0 & 3795 & 4554 & & 26.8 & 26.9 & & 0.8 & 1.0\\
F687W17  & 686.9 & 17.2 & 9270 & 12360 & & 27.2 & 27.1 & & 0.8 & 0.9\\
F704W17  & 703.7 & 17.9 & 6120 & 6120 & & 26.8 & 26.8 & & 0.9 & 0.9\\
\\ \hline \\
F806W17  & 806.5 & 16.1 & 14900 & 14900 & & 26.5 & 26.6 & & 0.9 & 1.0\\
F823W17  & 823.1 & 14.7 & 18570 & 24760 & & 26.8 & 26.8 & & 0.8 & 0.9\\
F840W17  & 840.0 & 15.6 & 19530 & 25872 & & 26.2 & 26.4 & & 0.9 & 0.9\\
\enddata
\tablecomments{(1) Filter name. (2) Central wavelength (in nm) of the filter for angle of incidence AOI=10.5$^\circ$ (approximately that for the center of the FOV). (3) Filter width (in nm). (4) Exposure time (in seconds). (5) Average 3-sigma depths (AB mag) for circular apertures of radius 0.8" for pointings 1 and 2. (6) Average seeing (in arcsec) for pointings 1 and 2.}
\end{deluxetable*}

GOODS-N is one of the most targeted areas of the sky at all wavelengths. For this work, apart from our new SHARDS data, we have combined the wealth of deep and high quality ancillary data, ranging from an ultra-deep X-ray exposure \citep[2Ms CDFN,][]{2003AJ....126..539A} to the deepest data in the MIR/FIR with surveys such as GOODS \citep{2004ApJ...600L..93G}, FIDEL \citep{2006ApJ...647L...9F}, PEP \citep{2011A&A...532A..90L}, HerMES \citep{2010A&A...518L..21O}, and Herschel-GOODS \citep{2011A&A...533A.119E}. Multiple spectroscopic redshifts for faint targets are also available \citep{2004AJ....127.3121W,2004AJ....127.3137C,2005ApJ...633..748R,2006ApJ...653.1004R,2008ApJ...689..687B,2011ApJS..193...14C,2015ApJS..218...15K}.

The SHARDS project was designed to be able to measure absorption indices such as the \Mg, or D(4000) for galaxies at {\it z}\,=1.0-2.5 through imaging data, and detect ELGs up to {\it z}\,$\sim$7. For those purposes, SHARDS obtained imaging data in the GOODS-N field through 25 medium-band filters covering the wavelength range between 500 and 950~nm in sub-arcsec seeing conditions. The typical width of our filters is 15 nm for wavelengths bluer than 880~nm, and 25--35~nm for the 3 reddest filters. In each filter, SHARDS is able to detect 26.5~mag sources at the 3$\sigma$ level (at least). Virtually all the deep region covered by GOODS with HST/ACS is surveyed by SHARDS using two GTC/OSIRIS pointings, summing up a total surveyed area of $\sim$130~arcmin$^2$ (see Figure~2 from \citealt{2013ApJ...762...46P}). The observations carried out by SHARDS allow to accurately determine the main properties of the stellar populations present in these galaxies through spectro-photometric data with a resolution R$\sim$50, sufficient to measure absorption indices such as the D4000 \citep[e.g.,][]{1983ApJ...273..105B,1999ApJ...527...54B,2003MNRAS.341...33K,2011ApJ...743..168K,2013MNRAS.434.2136H,2014MNRAS.443.3538H} or \Mg, index \citep{1997ApJ...484..581S,2004ApJ...614L...9M,2005MNRAS.357L..40S,2005ApJ...626..680D,2008A&A...482...21C,2013ApJ...762...46P}. At this spectral resolution, it is also possible to detect emission lines and measure their fluxes and equivalent widths. Among the lines SHARDS can detect, we can mention H$\alpha$, \OII, [OIII], Ly${\alpha}$, among others  (see e.g. \citealt{2011ApJ...740...47V,2013ApJ...762...46P,2013MNRAS.428.1128S}).

As explained in \citet{2013ApJ...762...46P}, given the special characteristics of the OSIRIS instrument at GTC, within each single frame taken with a given physical filter, each pixel sees a different passband. To overcome this issue, we performed a detailed calibration as a function of the position in the field of view (FOV). The significant variation of the passband seen by each point of the detector is a function of the position in the FOV and implies a complex behaviour of the absolute photometric calibration of the SHARDS images. To cope with these issues, we developed a special flux calibration procedure, aimed at determining the zeropoint of the SHARDS mosaics in each filter as a function of position in the image. The flux calibration of the SHARDS mosaics was performed by comparing the measured photometry in our images with spectroscopic data for several sources in the field.

To complement the SHARDS data, we also benefitted from the fact that GOODS-N has been observed by HST with ACS and WFC3 providing slitless, intermediate resolution spectroscopy in the optical (through the G800L grism; PEARS, \citealt{2009PASP..121...59K}; see also \citealt{2004ApJS..154..501P,2009ApJ...695.1591P}) and NIR (G102; PI: G.~Barro, G141; PI: B. ~Weiner). In addition, the availability of the deepest IRAC ([3.6 $\mu$m]$<$26.0~mag) and MIPS (F$_{5\sigma}$[24 $\mu$m]$>$30~$\mu$Jy) observations ensures the detection of the rest-frame NIR/MIR emission of the galaxies and allow us to estimate robust stellar masses and SFRs \citep{2005ApJ...630...82P,2008ApJ...675..234P,2008A&A...482...21C}.

We merged all these dataset for a mass-selected sample as described in \citet{2008ApJ...675..234P}. In that paper, we built spectral energy distributions (SEDs) using aperture photometry and deconvolution algorithms for IRAC bands. We then fitted the SEDs with a set of templates representative of the diverse galaxy populations and built with stellar population synthesis models. These fits allowed us to obtain estimations of the photometric redshifts and stellar masses. We describe in more detail this modeling in Sections~\ref{sect:rainbow} and \ref{sect:oii-prop}. Based on the models best-fitting the photometric data, we also obtained luminosities in the rest-frame UV (more specifically, at 150 and 280~nm), which were then converted to SFRs using the recipes from \citet{2011ApJS..193...30B}, taken from \citet{2005ApJ...625...23B}. Attenuation based on the UV slope $\beta$ were also estimated using the \cite{2000ApJ...533..682C} attenuation law and the SFR(IR)/SFR(UV) vs. $\beta$ calibration in Meurer et al. (1999). We will discuss in detail the UV-based SFRs in Section~\ref{sect:comp}. 

\subsection{Photometry and SED fitting}
\label{sect:rainbow}
In this section, we briefly summarise the basic steps and procedures adopted to extract the photometry and to fit the SEDs of different galaxy samples. These steps are performed exploiting  the {\rm Rainbow Cosmological Surveys Database}\footnote{The {\it Rainbow Cosmological Surveys database} is a vast compilation of photometric and spectroscopic data for several of the deepest cosmological fields, such as GOODS-North and South, COSMOS, or the Extended Groth Strip, among others. It is publicly accessible through the website: https:$//$rainbowx.fis.ucm.es$/$} hosted by the Universidad Complutense of Madrid, which is the central repository 
for all the SHARDS and ancillary data used in this paper.

Photometry is performed using the {\it Rainbow} (G.~Barro et al. 2015, in preparation) tools that are able to carry out simultaneous photometry  in multiple bands using prior based positions and apertures. The priors for the SHARDS extraction are based on SExtractor \citep{1996A&AS..117..393B} catalogs and \cite{1980ApJS...43..305K} apertures obtained as an average of those measured in all SHARDS bands. Photometry for longer wavelength bands (starting with IRAC) is extracted using circular apertures of fixed sizes.

For the SED fitting procedure, we adopt a two-population model, each population described by a SF history (SFH) following an exponentially declining law, characterized by timescales $\tau_{you}$ and a $\tau_{old}$ parameter for the {\it young} and the {\it old} stellar population, respectively. We use the models from \cite{2003MNRAS.344.1000B}, with a \cite{2003PASP..115..763C} IMF spanning stellar masses from 0.1 to 100~$M_\odot$. We assume the dust attenuation law from \cite{2000ApJ...533..682C}.
For the whole sample of \OII\ emitter candidates selected from SHARDS and the complementary  samples described in Sections~\ref{sect:datasets} and \ref{sect:complem} we derive stellar masses, SFRs, dust reddening A(V), and stellar ages from the best fits.

We compare the photometric data with a grid of models probing $\tau$ values in the range 6$\leq$log($\tau/yr$)$\leq$12 for each population.  These $\tau$ values were selected to include from instantaneous to roughly constant SFHs.  The grid of models spans a range of ages, from 1~Myr to up to 1~Gyr for the young population and from 1~Gyr  up to the age of the Universe at the given galaxy redshift for the old population. We adopt a \cite{2000ApJ...533..682C} dust attenuation law and dust extinctions with A$_{you}$(V)$=$0.0--7.0~mag and A$_{old}$(V)$=$0.0--2.0~mag, with 0.1 mag increments. We consider the six metallicities available from the \cite{2003MNRAS.344.1000B} libraries. Additional details and an example of the stellar population fits in the SHARDS spectral region are given in \ref{sect:linflux}. In these fits, the redshifts are fixed to either the spectroscopic values when available or the best photometric redshift solution. As we show in Section~\ref{sect:oiisel}, we have a considerable improvement on the photometric redshift determination using the full SHARDS dataset to perform SED fitting (G.~Barro et al. 2015, in preparation; see also \citealt{2014MNRAS.444..906F}).

We remark that the results obtained from this modeling are of particular relevance for the determination of the continuum at the wavelength of \OII\ emission line, as discussed in Section~\ref{sect:oii-prop}.

\begin{figure*}[tb]
  \begin{center}
    \includegraphics[width=0.49\textwidth]{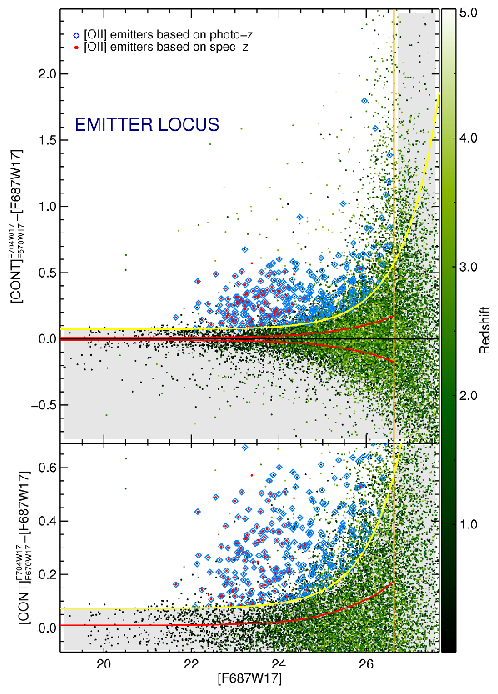}\hspace{0.2cm}
    \includegraphics[width=0.49\textwidth]{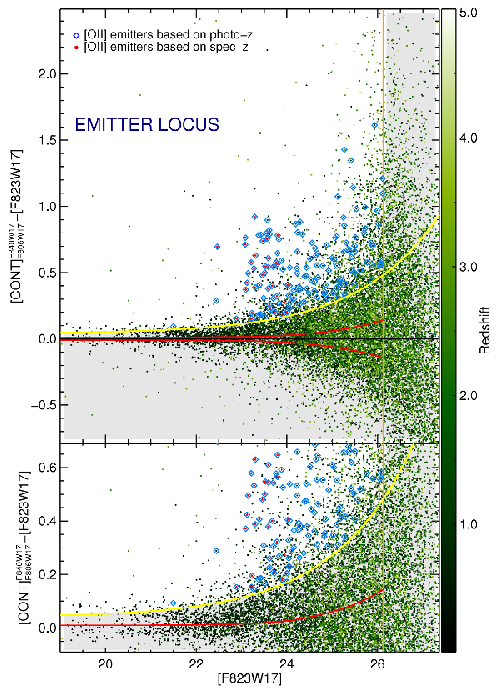}
    \figcaption{\label{fig:trumpet} Color-magnitude diagram showing ELG 
      candidates at $\sim$687~nm ({\it left
      panel}) and $\sim$823~nm ({\it right panel}). The vertical
      axis shows the color between a central SHARDS filter band and
      the average magnitude in the adjacent (continuum) SHARDS bands.
      The locus for galaxies
      with an emission line within the filter width detected with more
      than 2.5$\sigma$ confidence is the region above the yellow-dashed curve,
      identified as the {\it emitter locus}.  The vertical shaded orange
      line shows the minimum 3$\sigma$ detection threshold of the
      SHARDS survey in the selected filter bands. Red dots indicate
      \OII\ emitters candidates with spectroscopic confirmation, 
      while blue diamonds stand for
      candidates selected using photometric redshifts. The red-dashed lines
      depict the typical photometric uncertainty as a function of the filter magnitude.
      Lower panels are a zoomed view around the selection curve.}
  \end{center}
\end{figure*}

\subsection{Sample selection of \OII\ emitters}
\label{sect:datasets}
We have developed a novel selection technique to identify ELGs using the SHARDS medium-band spectro-photometric dataset, based on similar narrow-band selection techniques. The technique to select star-forming galaxies (and also AGNs) from medium-band photometry is based on comparing the flux measured in one filter (the \emph{central filter}, hereafter) with those obtained in two adjacent filters (the \emph{continuum filters}, hereafter). Galaxies presenting emission lines would then pop-up in the central filter if their redshifts move the line within the passband of this filter, thus providing the emission line flux, while the adjacent filters would give us an estimate of the continuum around the emission line. 

A similar color-excess technique has been applied to multiple surveys of intermediate, high, and very high redshifts (focusing on different lines such as H$\alpha$ or Ly$\alpha$) using typically narrow-band passbands as the central filter and a broad-band passband as the continuum filter (see e.g. \citealt{2008ApJS..176..301O,2008ApJ...677..169V,2011ApJ...740...47V,2009MNRAS.398L..68S,2009MNRAS.398...75S,2012MNRAS.420.1926S,2013MNRAS.428.1128S,2014MNRAS.440.2375M}).

Note that our dataset consists of medium-band filters, not narrow-band, with a typical equivalent spectral resolution $R\sim50$ to be compared, e.g., to $R\sim100$ in \cite{2008ApJ...677..169V,2011ApJ...740...47V} or $R\sim80$ in \cite{2009MNRAS.398...75S,2012MNRAS.420.1926S,2013MNRAS.428.1128S}, but the depth and image quality of the SHARDS images, jointly with the use of adjacent medium-band passbands for the determination of the continuum, offer several advantages over previous surveys as detailed next. Note also that the SHARDS spectral resolution is similar to that achieved with HST optical grism spectroscopy.

\subsubsection{Selection of ELG candidates using SHARDS data}
\label{sect:elgsel}
\begin{figure*}[tb]
  \begin{center}
    \includegraphics[width=0.49\textwidth]{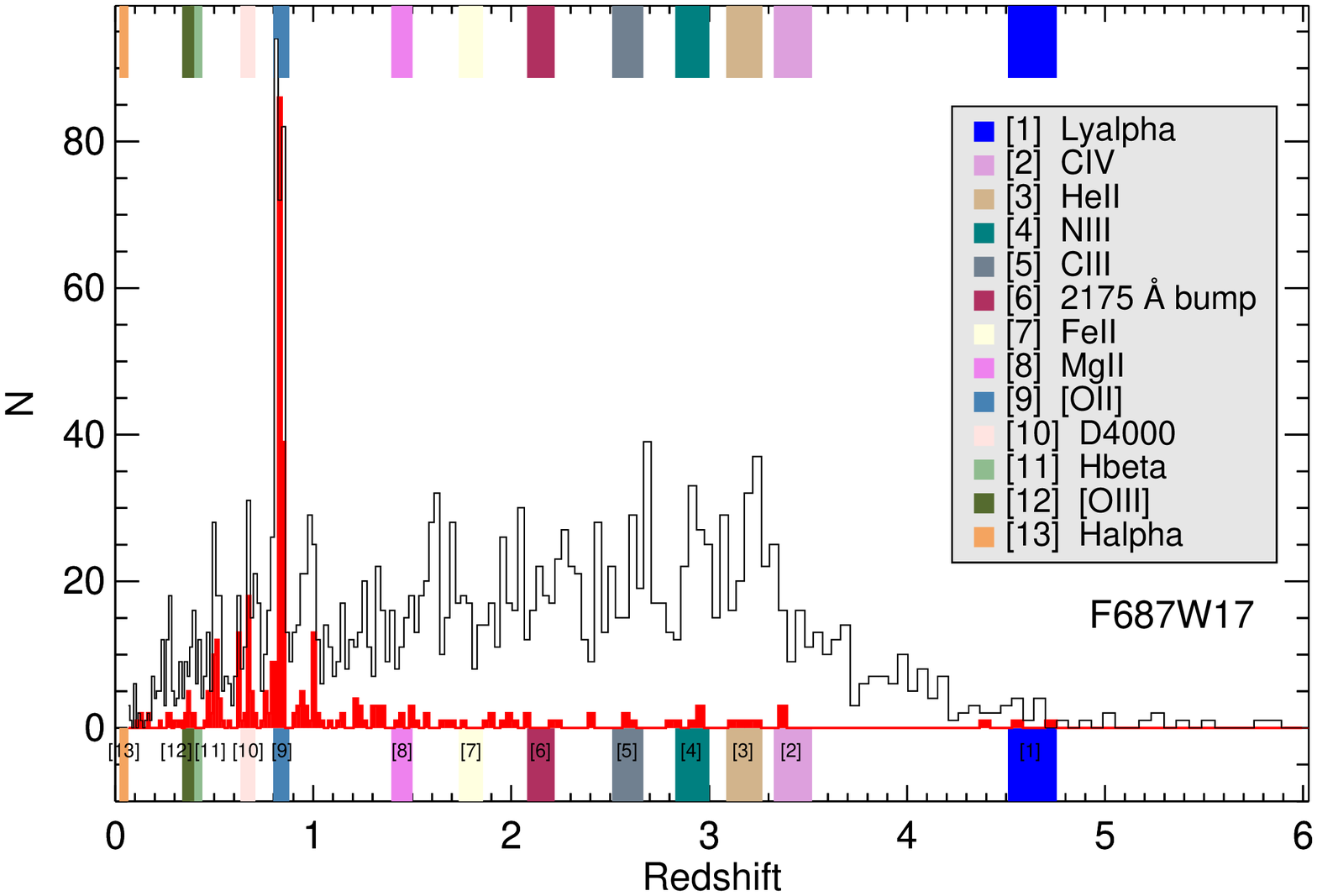}
    \includegraphics[width=0.49\textwidth]{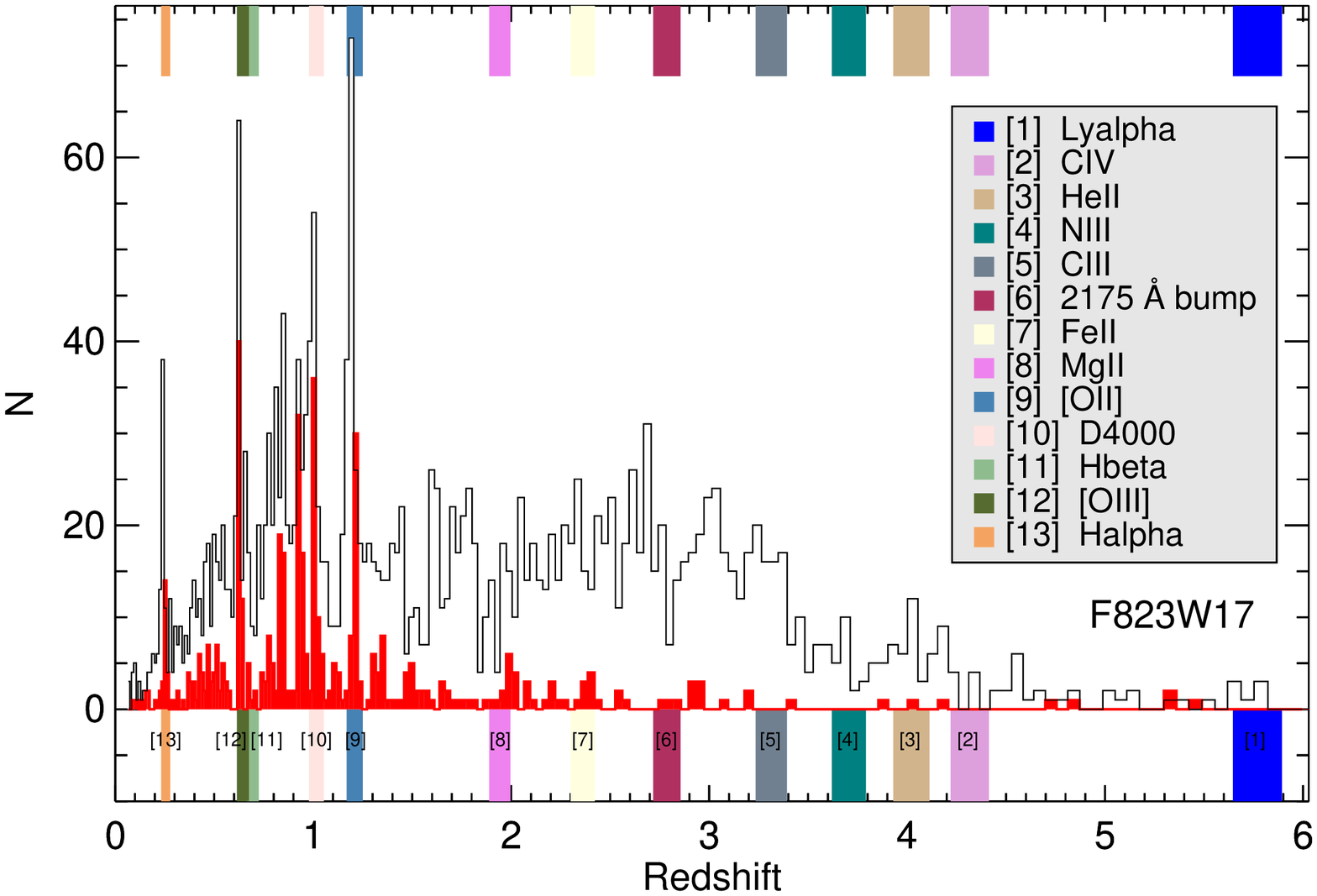}
    \figcaption{\label{fig:distr} Spectroscopic and photometric
      redshift distributions for the sources selected as ELGs
      candidates with the F687W17 and F823W17 filters. The histograms
      in red refer to galaxies with spectroscopic redshift
      confirmation. The outlined black histograms refer to
      photometric redshifts (\citealt{2008ApJ...675..234P}; G.~Barro et al.
      in preparation). The expected redshifts for emitters with some
      of the most typical lines (e.g., Ly$\alpha$, \OII, [OIII], or
      H$\alpha$) lying in each one of the two central filters are 
      marked with shaded regions according to the legend of each panel. 
      Numbers below the histogram distributions help to recognise each line. 
      We also mark other spectral features such as the Mg-Fe absorption band at
      $\sim$280~nm, or the 2175~\AA\, dust absorption bump,  
      which may mimic emission lines at a different redshift. Bin sizes
      are proportional to $\Delta(z)/(1+z)$$\sim$0.006,
      our typical photometric redshift accuracy.}
  \end{center}
\end{figure*}
In this work, we used two SHARDS filters as central passbands, F687W17 and F823W17, two of the deepest in our survey, centered at 687 and 823~nm, respectively.  The most common emission line detected with these filters is [OII]$\lambda$3727, which means that the central wavelength {\ bf corresponds} approximatively to a redshift of {\it z}\,$\sim$0.84 and {\it z}\,$\sim$1.23 for the bulk of our emission-line galaxy sample. We start the analysis of SHARDS selected ELGs with these two filters because (1) they are representative enough of the whole dataset, (2) they are two of the deepest ones, and (3) they enable the study of samples of galaxies located at sufficiently separated redshifts in order to unveil the possible evolutionary effects on their derived physical properties. In a future work we will present more extensive results including all the SHARDS filters.

In order to be able to compute the continuum level, we have also used data in four additional filters (lying on the bluer and redder side of
each central filter), namely: F670W17, F704W17, F806W17, and F840W17. The main characteristics of the central and continuum filters and observed data used in this work are summarized in Table~\ref{table:filters}.

A sketch of the technique used in the selection of the ELGs sample is presented in Figure~\ref{fig:trumpet}. We compare the magnitude in the central filter (F687W17 or F823W17) with the continuum filters (a combination of the 2 adjacent ones for each central filter).
Typically, the estimation of the continuum in ELGs is carried out with broad-band filters, which also includes the line entering in the central filter and possibly others \cite{2008ApJ...677..169V,2011ApJ...740...47V,2009MNRAS.398...75S,2012MNRAS.420.1926S,2013MNRAS.428.1128S}. By using two adjacent medium-band filters, we can estimate a more accurate continuum level. This method can be considered robust for the continuum determination as far as the contribution of the emission line to the side filters can be neglected, which is typically the case for SHARDS filters (given the filter shape). Indeed, the SHARDS continuum coverage of the optical spectral range ensures that intrinsic variations of the spectral slope do not affect the continuum determination. The probability of another line entering in the continuum filter is also smaller than when using broad-band filters. Nonetheless, the finite width of the filters, the noise, and other spectral features possibly falling in our observed spectral range can affect the final measurement. We rely on the interpolation between the two filters to define the observed continuum for the selection of ELGs.  However, once the emission line candidates are selected, we perform a detailed SED fitting to fine-tune the continuum determination in order to obtain robust line fluxes and equivalent widths (as described in Sections~\ref{sect:rainbow} and  \ref{sect:linflux}).

Figure~\ref{fig:trumpet} shows the selection diagrams for the filters F687W17 (left panel) and F823W17 (right panel). They depict the
"continuum$-$line color" vs. the "central-filter magnitude" for each of the two filters. As we move to fainter magnitudes, photometric errors become larger and larger, resulting on a typical ``trumpet'' shape of the cloud of points. The color vs. magnitude distribution can be used to estimate the typical scatter as a function of magnitude. In this plot, $\sigma$ is the standard deviation of the colors, and we calculate it as a function of magnitude. Galaxies with significantly larger colors than the statistical standard deviation (associated to the photometric uncertainties) are identified as sources with an excess of light in the central filter, i.e., emission-line galaxy candidates. We have chosen a selection function (yellow curve) coming from a 2.5$\sigma$ cut in the color-magnitude distribution. The vertical orange line represents the minimum detection threshold adopted for the selection, corresponding to the minimum 3$\sigma$ detection limit of the three filters used in each case (selection band plus the two contiguous filters, see Table~\ref{table:filters}).  
These cuts, provide a good compromise between the inclusion of the largest number of ELG candidates and the contamination from spurious sources (see, e.g., \citealt{2008ApJ...677..169V,2011ApJ...740...47V}). Summarizing, galaxies identified as potential emitters
are found in the {\it emitters locus} region of these diagrams. Note that the selected galaxies are candidates to have an emission line located in the central filter. This line could be, for the F687W17 filter, \OII\ at {\it z}\,$\sim$0.84, but it may also be any other line such as \lya at {\it z}\,$\sim$6 or \hal at {\it z}\,$\sim$0, among others.

The redshift distributions of the galaxies found in the emitter loci presented in Figure~\ref{fig:trumpet} are shown in Figure~\ref{fig:distr}. Most of the peaks in the distribution are actually due to an emission line falling within the central filter observational window. Spikes can also be produced by absorption lines and/or breaks in the continuum falling in one of the filters used for the continuum definition, which can mimic an emission line at a different redshift (see Fig.~9 in \citealt{2013ApJ...762...46P}).
For example, a number of galaxies can be selected due to the D4000 break \cite[see][]{2013MNRAS.434.2136H} falling in the filter used to define the left side of the continuum, mimicking an emission line at lower redshift. 

\begin{figure}
  \begin{center}
    \includegraphics[width=1.05\columnwidth]{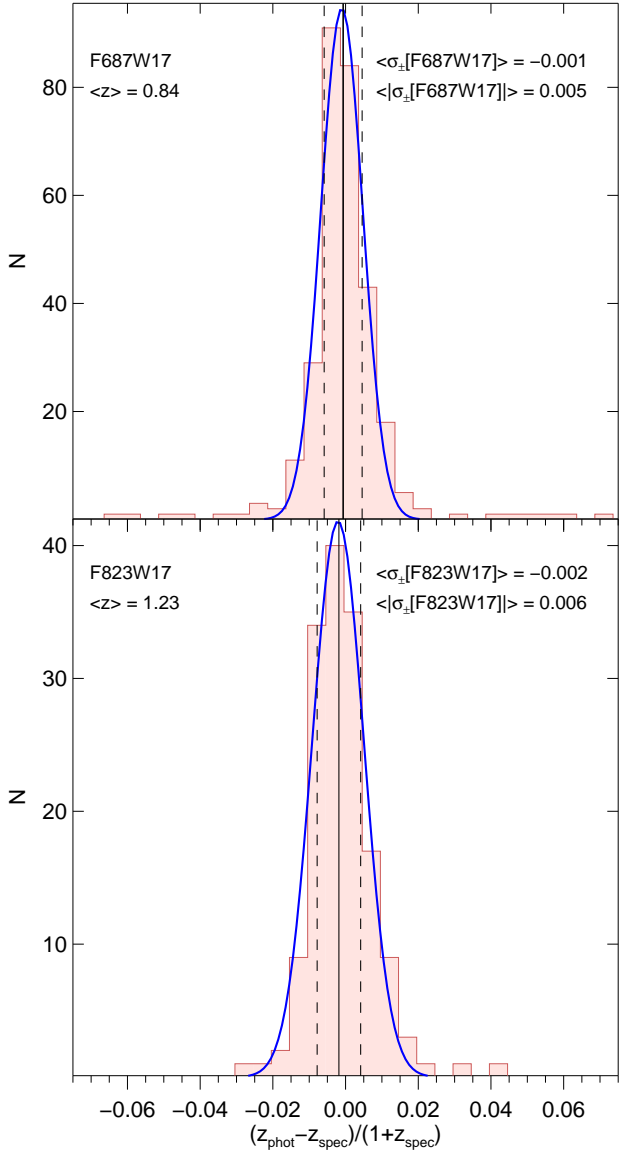}
    \figcaption{\label{fig:zcheck}Comparison between spectroscopic and
      photometric redshifts for SHARDS sources around {\it z}\,$\sim$0.84
      ({\it top panel}) and {\it z}\,$\sim$1.23 ({\it bottom panel}), the two
      redshifts where we present \OII\ emitters selected with the
      F687W17 and F823W17 filters.  We show the distributions of
      redshift accuracies,
      $\sigma_{\pm}=(z_{phot}-z_{spec})/(1+z_{spec})$. In both panels,
      the blue curve is the Gaussian fit to the distribution, with the 
      vertical solid and dashed lines representing the mean offset
      and 1$\sigma$ dispersion respectively.}
  \end{center}
\end{figure}

The photometric redshift distribution broadly follows the shape outlined by the spectroscopic sample, revealing their high quality. In Figure~\ref{fig:zcheck}, we show a comparison between the SHARDS-derived photometric redshifts and available spectroscopic redshifts in the GOODS-N field for our two redshift ranges around {\it z}\,$\sim$0.84 and {\it z}\,$\sim$1.23. We have estimated a typical accuracy of $\Delta(z)/(1+z)$$=$0.006 by comparison of the photometric redshift with their corresponding spectroscopic measurements at all redshifts and magnitudes. In the intervals covered by the two filters used in this work, we derived a typical rms $\sim$0.5\%. The high quality of the photometric redshifts is very relevant when aiming at selecting a particular class of ELGs, such as \OII\ emitters at two
different redshifts, in our case. These photometric redshifts will be presented in another work (G.~Barro et al. 2015, in preparation).

For this study, we focus only on the \OII\ emitters selected combining Figures~\ref{fig:trumpet} and \ref{fig:distr}. To make the most robust selection, the tolerance allowed for the inclusion of \OII\ ELG candidates is different depending on the type of redshift estimate at hand. Galaxies with already available spectroscopy are easily confirmed as \OII\ emitters selected by SHARDS. The selection of galaxies with no spectroscopic confirmation must be done with photometric redshifts, accounting for their uncertainty (see the next section for more details).  

\subsubsection{Identification of \OII\ emitters candidates}
\label{sect:oiisel}
\begin{figure*}[tb]
  \begin{center}
    \includegraphics[width=1.\textwidth]{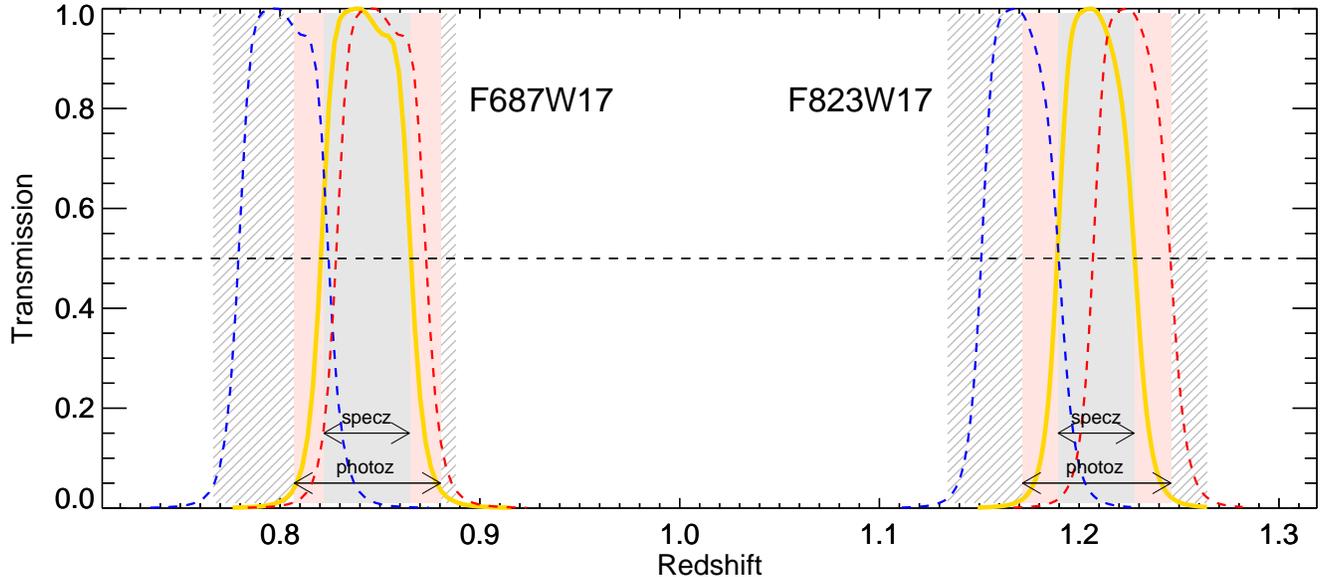}
    \figcaption{\label{fig:filters}Scheme of the redshift dependence
      of the selection of \OII\ ELGs based on the
      SHARDS medium-band data. We show the results for filters F687W17
      ({\it z}\,$\sim$0.84) and F823W17 ({\it z}\,$\sim$1.23). Blue (red) dashed
      lines show the transmission curves for the bluest (reddest)
      passbands seen by any galaxy in the SHARDS images for these filters. The orange
      curves represent the nominal central passband for the given
      SHARDS filter (centered on 687 and 823~nm).  The filled grey area
      represents the interval where galaxies with spectroscopic
      redshifts are selected. The shaded pink area indicates the
      redshift interval of the selection for galaxies with photometric
      redshifts. Its width is based on the photometric redshift mean
      error $<\delta({\it z})>=\sigma\times(1+{\it z})$ at redshift $z$, where
      $\sigma=0.005$ for {\it z}\,$\sim$0.84 and $\sigma=0.006$ for
      {\it z}\,$\sim$1.23. The 50\% normalized transmission is
      marked with a black dashed lines. The grey-hatched area gives the 
      global envelope of the transmission curves along the FOV.}
  \end{center}
\end{figure*}

Therefore, the question is how we can identify \OII\ emitters and separate them from galaxies with other emission lines or bumps. 
We identify {\it bona fide} [OII] emitters as those ELG candidates for which the spectroscopic redshift is such that the  \OII\ line 
lies within the interval implicitly defined by the FWHM of the observed central passband. Note that, as explained in Section~\ref{sect:datasets} and \citet{2013ApJ...762...46P}, the actual passband seen by each galaxy depends on the exact position in the GTC/OSIRIS FOV. The procedure is graphically explained in Figure~\ref{fig:filters}. The medium-band technique is very effective in selecting ELGs in a narrow range of redshifts. If we had spectroscopic confirmation for all sources, those ranges would be as narrow as $\Delta(z)$$\sim$0.05. However, $\sim 53\%$ of of galaxies finally identified as \OII\ emitters at {\it z}\,$\sim$0.84 and $\sim74\%$ of the sample {\it z}\,$\sim$1.23, only count on a photometric redshift.  In order to determine which galaxies are actual \OII\ emitters on the basis of photometric redshifts, we have to consider the error of those photo-{\it z}'s. 

In addition to this, due to the variable central wavelength (CWL) along the FOV of the OSIRIS instrument, we obtain a continuum distribution for the selection function that depends on the CWL of the filter for each galaxy position. Combining all effects, we obtain the hatched region in Figure~\ref{fig:filters}, which represents the overall extension in the redshift space for our selection function based on
the {\it trumpet} plots in Figure~\ref{fig:trumpet} and the photometric redshifts (see Figure~\ref{fig:distr}). We remark that this redshift range does not apply to the selection in the whole area surveyed by SHARDS, the redshift range probed actually depends on the
position of the galaxy in the FOV.

This procedure provides two catalogues of ELGs selected as \OII\ emitters based on the two redshift estimates: one using photometric
redshifts, and one on the basis of spectroscopic redshifts, which should be a subsample of the former. We call the latter the specz-confirmed ELGs, while the former is the photoz-selected sample. These catalogues can be merged to get a robust and complete selection.  
Adopting a similar procedure, we are able to produce catalogs for each possible detected line (as those listed in the legend of Fig.~\ref{fig:distr}) and we plan to analyze the full combined set of detected lines in a future work.

\begin{figure*}[tb]
  \begin{center}
    \includegraphics[width=.99\columnwidth]{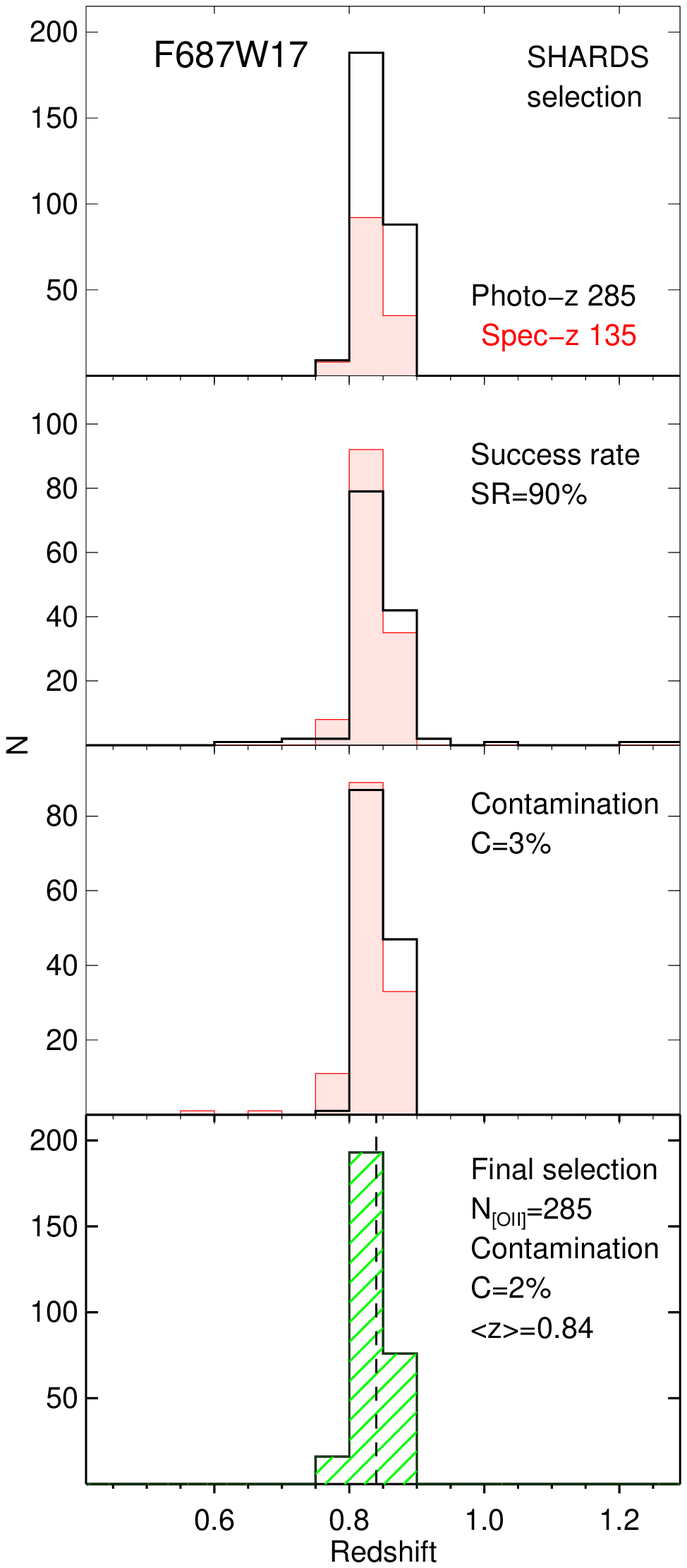}
    \includegraphics[width=.99\columnwidth]{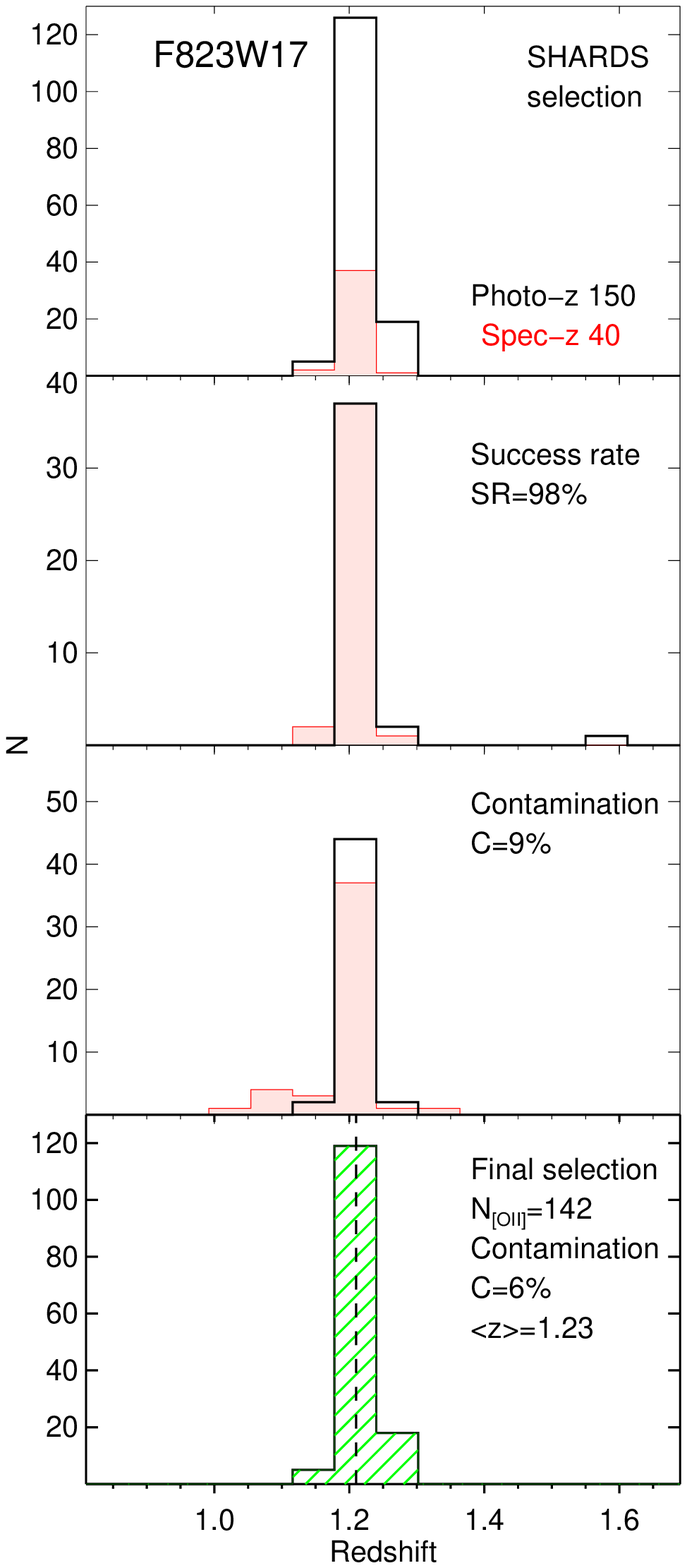}    
    \figcaption{\label{fig:fighist} Redshift histograms showing the
      selection, success rate and contamination rate for the \OII\ ELG
      candidate samples built with central filters F687W17 ({\it z}\,$\sim$0.84, plots
      on the left) and F823W17 ({\it z}\,$\sim$1.23, right plots) using
      spectroscopic or photometric redshifts. {\it Top panels:}
      redshift distributions for the selected objects using
      spectroscopic (red filled histogram) or photometric (black
      histogram) redshifts.  {\it Second row of panels:} redshift distribution
      of objects selected on the basis of their spectroscopic redshift (red filled
      histogram) compared to the photometric redshift distribution for
      the same galaxies (black histogram). The ratio in the numbers
      of objects in these distributions defines the {\it success
        rate} (SR).  {\it Third row of panels:} redshift distribution for objects
      selected on the basis of their photometric redshift (black histogram)
      compared to the distribution of their corresponding
      spectroscopic redshifts (red filled histogram) for those objects in which 
      both redshift estimates are available. The normalised
      difference in the numbers of objects in these distributions defines
      the {\it contamination} (C).  {\it Bottom panels:} redshift
      distribution for the final selected sample built by complementing the
      spectroscopic and photometric redshift selections. 
      The vertical dashed line represents the median redshift in each panel.}
  \end{center}
\end{figure*}

In order to assess and improve the selection of \OII\ emitters, we compare the spectroscopic and photometric selected samples in Figure~\ref{fig:fighist}. In the top panels, we plot the redshift distribution of the two samples of \OII\ emitters based on spectroscopic and photometric redshifts. The photometric redshift selection closely follows the spectroscopic distribution, but it includes a higher number of galaxies. If the galaxies selected with photometric redshifts are {\it bona fide} \OII\ emitters, then half of the {\it z}\,$\sim$0.84 sample and one fourth of the {\it z}\,$\sim$1.23 sample currently would have a spectroscopic confirmation (i.e., the spectroscopic completeness is $\sim$50\% and $\sim$25\%).

It is interesting to note that the galaxies with no spectroscopic confirmation are typically fainter than AB$\sim24.5$, as clearly shown
in Figure~\ref{fig:trumpet}. In fact, the spectroscopic completeness for the selected {\it bona fide} [OII] emitters brighter than $R=24.5$ is $\sim91\%$ $(74\%)$ in the F687W17 (F823W17) filter, and $\sim 13\%$ $(16\%)$ for galaxies with $R$$>$24.5~mag.  This is the spectroscopic limit of the redshift surveys carried out in the GOODS-N field, and the typical detection threshold for the vast majority of data taken with state-of-the art spectrographs in 10-meter class (or smaller) telescopes. The SHARDS observations reach at least 2 magnitudes fainter, and thus open the possibility to reliably select and study fainter and/or higher redshift ELGs. We note, furthermore, that the number of galaxies having publicly available spectra of good quality, where the $S/N$ is sufficiently high to derive a robust equivalent width (EW) and \OII\ line flux determination, only represents a small fraction (typically $\sim5-10\%$) of the total number of galaxies for which a spectroscopic redshift determination is available in this study.

In the second and third rows of Figure~\ref{fig:fighist}, we compare more quantitatively the samples selected with spectroscopic and
photometric redshifts in terms of the success rate and the contamination. The success rate is defined as:

\begin{equation}
SR=N_{ph-conf}/N_{sp} 
\end{equation}

\noindent where $N_{sp}$ is the number of galaxies in the spectroscopic catalog and $N_{ph-conf}$ is number of galaxies with a
spectroscopic confirmation which are included in the photometric redshift sample. This number is related to the ability to select a galaxy using the photometric redshift and provides an estimate of the number of galaxies we lose due to uncertainties in the photo-z or a failure in the detection of the emission-line with the SHARDS filter. Typically, the selection based on SHARDS medium-band data and photometric redshifts has a success rate higher than 90\%.

In an analogous way, we define the contamination as:

\begin{equation}
C=1-N_{ph-conf}/N_{ph-sp} 
\end{equation}

\noindent where $N_{ph-sp}$
is the total number of galaxies in the photometric catalog having a spectroscopic redshift estimate (either within the expected redshift range or outside). This number gives an estimate of the fraction of contaminants expected in the final sample. The contamination measured in both redshifts bins considered in our study is less than 10\%. The measured success rate and contamination level are coherent with the expectations due to the photometric redshift errors and spectroscopic uncertainties for the faintest observed galaxies.

\begin{figure*}[tb]
  \begin{center}
    \includegraphics[width=0.49\textwidth]{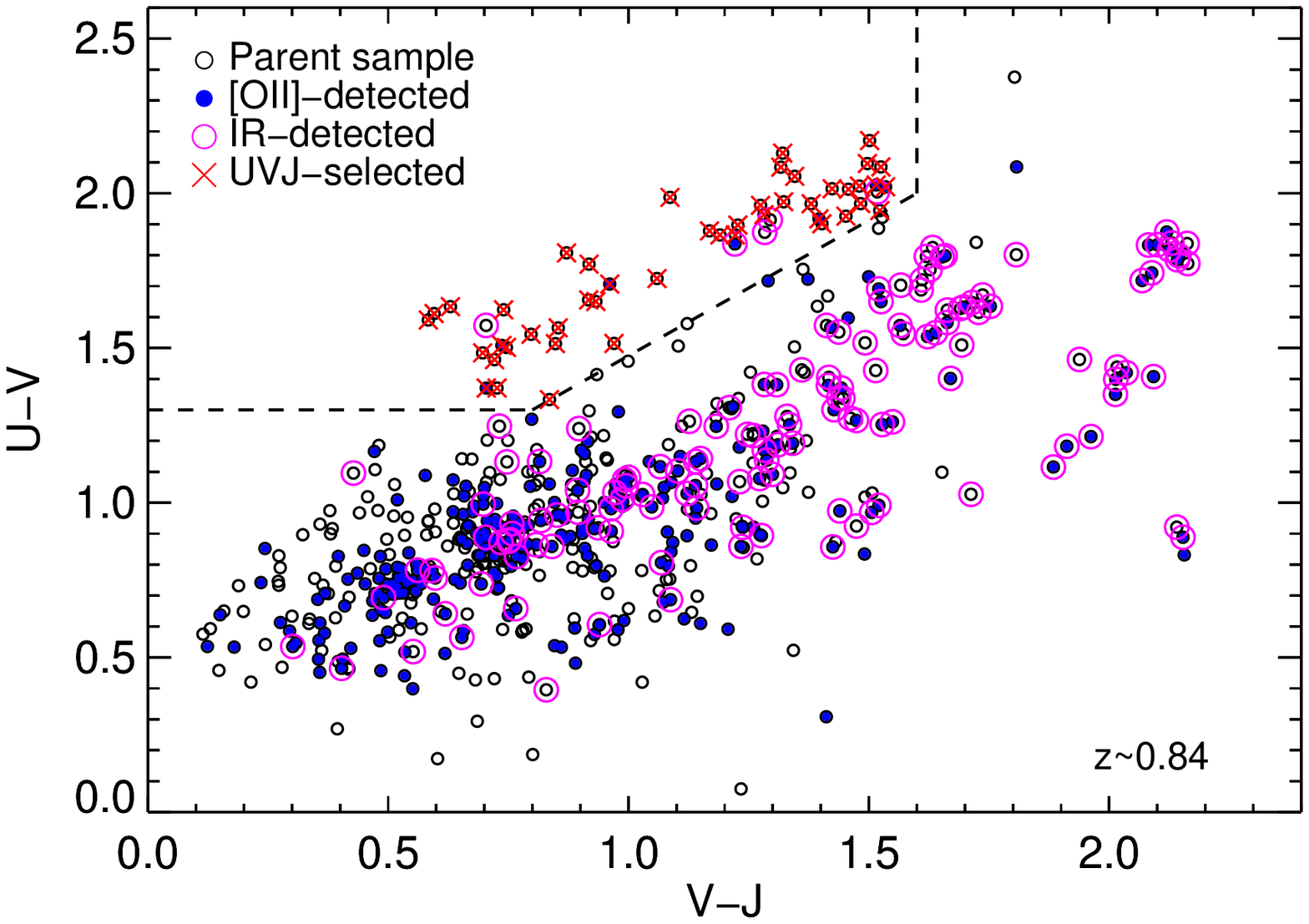}
    \includegraphics[width=0.49\textwidth]{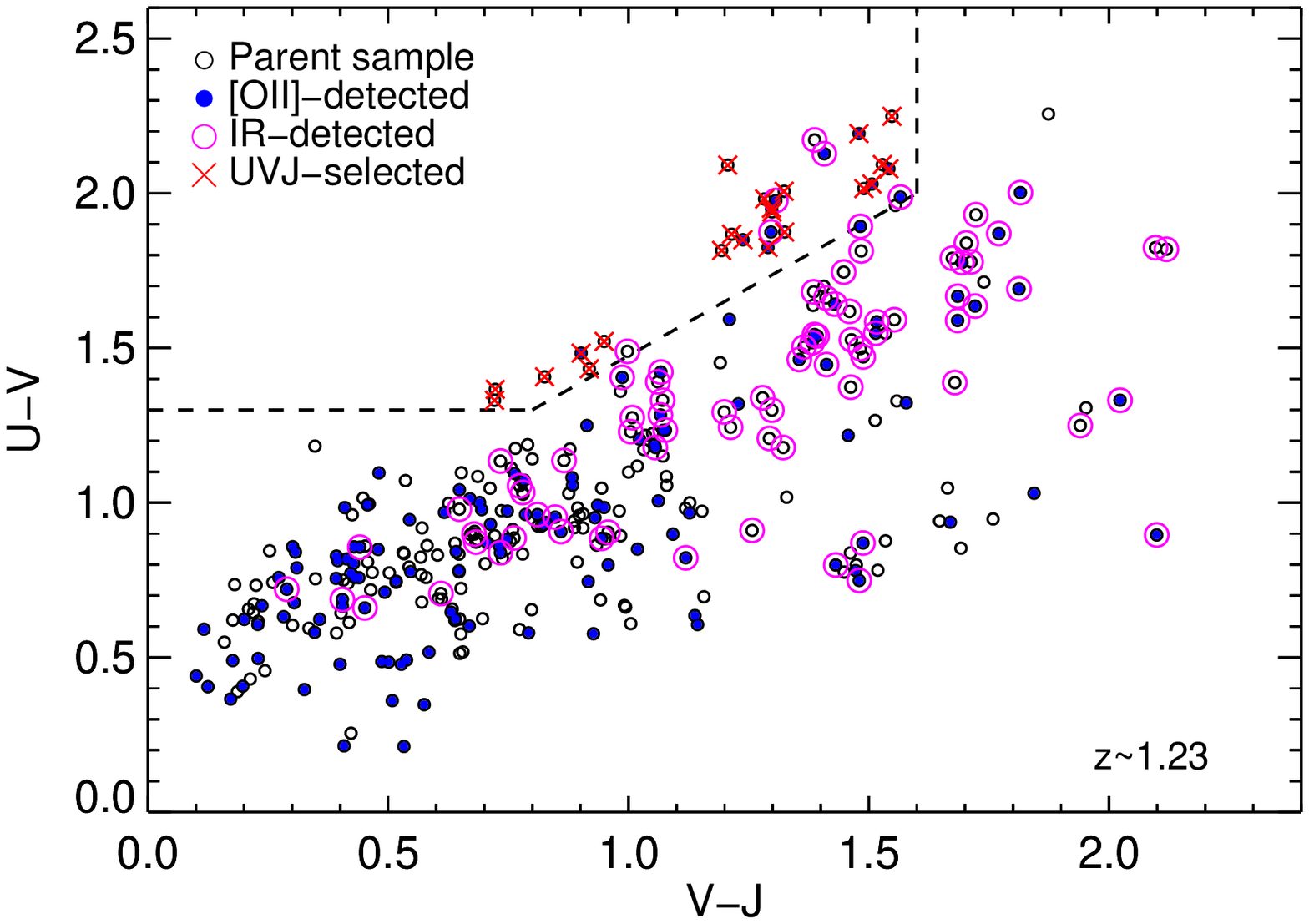}  
    \includegraphics[width=0.49\textwidth]{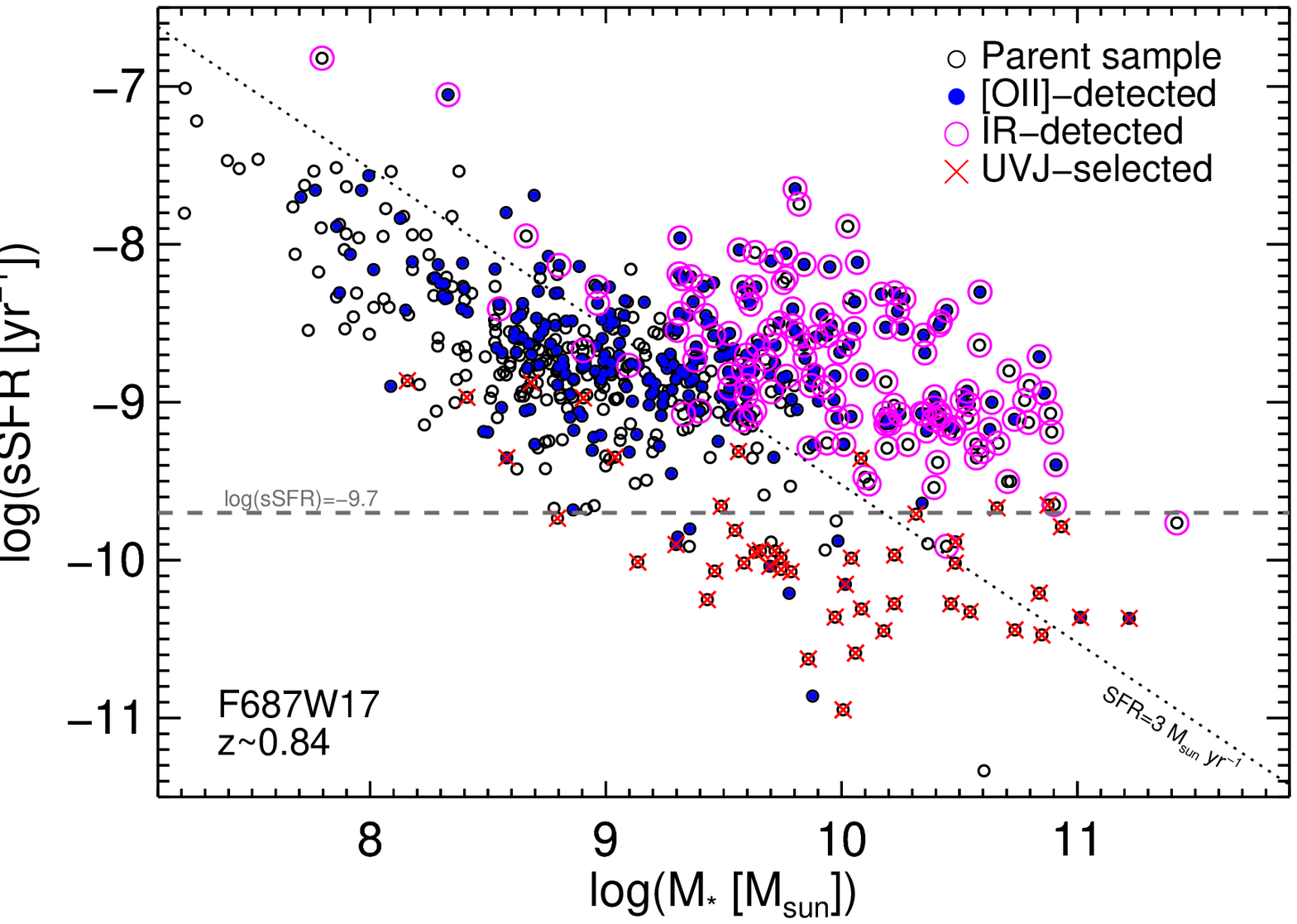}
    \includegraphics[width=0.49\textwidth]{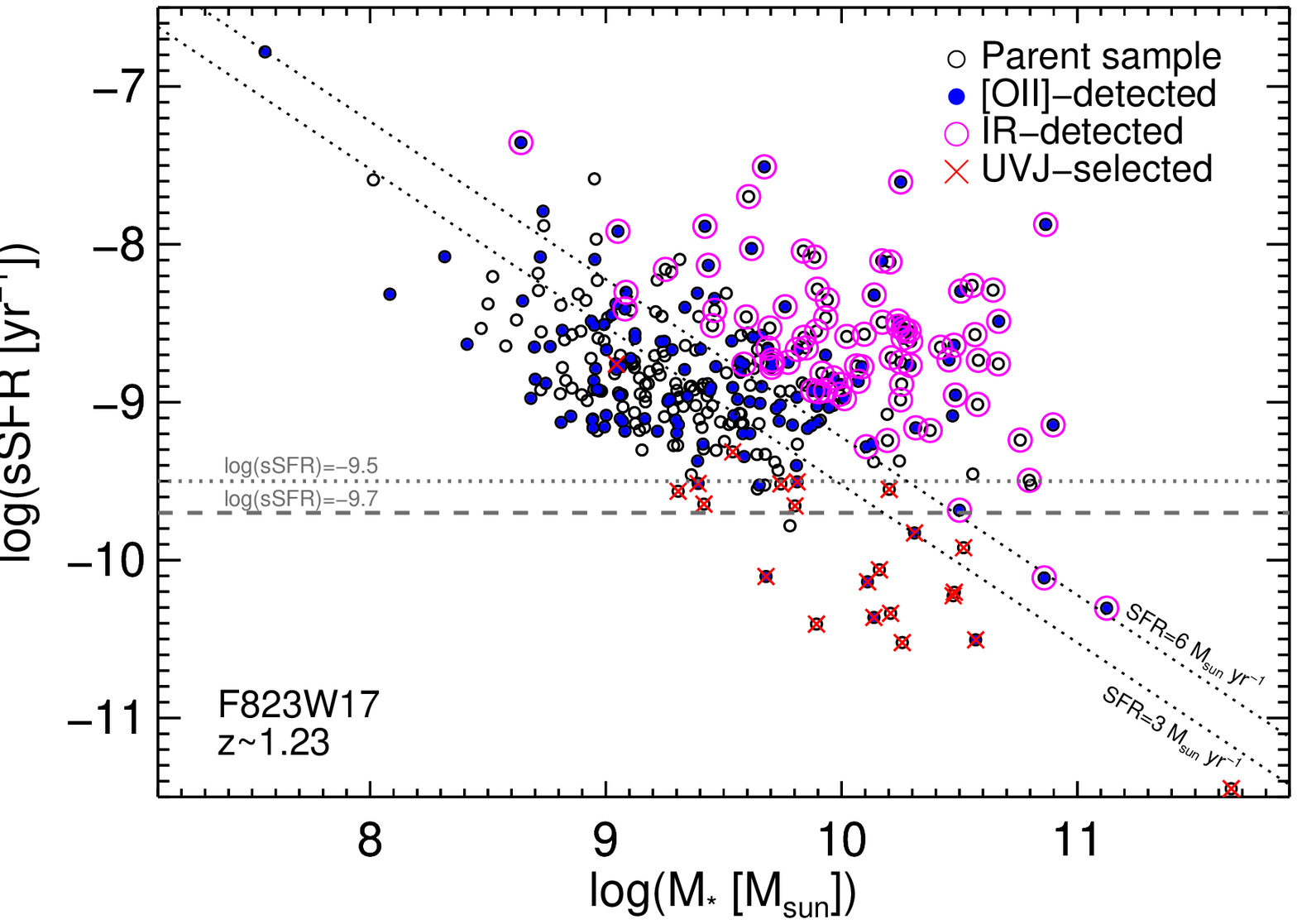}
    \figcaption{\label{fig:nf0} {\it Top panels: }UVJ color-color diagram with the quiescent galaxies region defined as in Whitaker et al. 2011 (F687W17 to the left, F823W17 to the right). IR detected galaxies within the UVJ quiescent region are excluded from the passive sample. The different samples are indicated with different symbols identified in the legend. {\it Bottom panels: } sSFR-M$_*$ diagrams for the two SHARDS filter selections at z$\sim0.84$ (left panel) and z$\sim1.23$ (right panel). The different subsamples are plotted with different symbols identified in the legend. UVJ passive galaxies are defined as in \citealt{2011ApJ...735...86W}. The dark-grey dashed and dotted horizontal lines mark two cuts at constant sSFR (log(sSFR [yr$^{-1}$])=$-$9.7 and $-$9.5), adopted for the sSFR-based definition of quiescent galaxies. The black-dotted lines mark the sSFR-M$_*$ relations for constant values of SFRs (SFR=3 and 6~M$_{\rm sun}$~yr$^{-1}$), corresponding approximatively to our detection limits for IR-selected galaxies in the two redshift ranges.}
  \end{center}
\end{figure*}

In the bottom panels of Figure~\ref{fig:fighist}, we show the redshift distribution for the final sample of {\it bona fide} \OII\ emitters
built by merging the samples selected on the basis of spectroscopic and photometric redshifts. These final catalogs do not include the known contaminant galaxies identified on the basis of measured spectroscopic redshifts. Accounting for this, the final contamination fractions are 2\% and 6$\%$ for the filters F687W17 and F823W17, respectively. Obviously, both the contamination and the success rate for the final selection are very sensitive to the photometric redshift errors. We have verified the effect of using photometric redshift catalogs derived using only broad-band photometry (that is, excluding SHARDS data), with an accuracy of $\Delta(z)/(1+z)$$\sim$0.03, typical of most photometric redshift catalogs.  The success rate decreases to $\sim70\%$ level, and the contamination increases to $\sim40\%$. Henceforth, we will always refer to the sample shown in the bottom panels of Figure~\ref{fig:fighist} as the {\it final \OII\ emitters sample}.

\subsection{Complementary samples: mass-selected, IR-detected and quiescent galaxies}
\label{sect:complem}
As already mentioned before, we use in our analysis the ancillary multi-wavelength catalog and advanced products in the GOODS-N field presented in \cite{2008ApJ...675..234P} and compiled in the Rainbow Cosmological Surveys Database \citep[see,][]{2008ApJ...675..234P,2011ApJS..193...13B}. This dataset includes observations from X-rays to the Far-IR and radio bands, as well as spectroscopic data in the GOODS-N field from the literature. In \cite{2008ApJ...675..234P}, we presented a merged photometric catalog including broad-band data for a stellar mass selected sample based on ultra-deep IRAC observations. For this work, we have merged this catalog with the SHARD dataset. As described in \cite{2008ApJ...675..234P},  the parent sample used in this paper has been built using IRAC luminosities as proxies for the stellar mass. Thanks to the ultra-deep IRAC imaging available our sample is mass-complete for galaxies with M$>$10$^9$M$_{sun}$ up to redshift z$=$1.

In order to perform a comprehensive study of SFGs at {\it z}\,$\sim$0.84 and {\it z}\,$\sim$1.23 , we have built a general parent sample, starting from the GOODS-N mass-selected sample, by selecting all the galaxies that would enter the medium-band filter selection using the SHARDS data (i.e. with the same redshift distribution as our \OII\ emitters), independently from their emission properties. 

We apply the same selection procedure presented in Section~\ref{sect:oiisel} for the selection of \OII\ emitters, but this time relaxing the condition on the emission properties of the galaxies. This means that the position-dependent redshift selection function (as schematically represented in Figure~\ref{fig:filters}) will not be applied only to the ELG candidates (i.e. galaxies within the {\it emitter locus} in Figure~\ref{fig:trumpet}) but to the whole mass-selected galaxy sample, independently from the emission properties. In other words,  we check the position within the FOV of each galaxy selected from its (photometric or spectroscopic) redshift, and then determine whether that redshift is good for a selection based on the SHARDS filters should the galaxy had had an emission line. In this way we can define a parent sample that includes and complements our \OII\ emitter compilation and provides a complete census of galaxies (both star-forming and also quiescent) at those redshifts, which we simply dub the {\it parent sample} hereafter.

Starting from this general selection, we can derive different subsamples according to several indicators for the SFR. In fact, we can estimate the SFR from the rest-frame UV luminosity (corrected for attenuation) derived from the SED fitting for all the parent sample, 
then we can define the subsamples of [OII]-detected (according to our SHARDS selection) and IR-detected galaxies (for example, using Spitzer/MIPS 24~$\mu$m data). Finally, we  can define a sample of {quiescent galaxies} using the rest-frame (V$-$J) versus (U$-$V) color-color diagrams following \cite{2011ApJ...735...86W}, or alternatively as all galaxies lying below a certain sSFR cut (we use 0.2~Gyr$^{-1}$ as our arbitrary limit) and not presenting IR emission.

Interestingly, the sample of quiescent galaxies defined on the basis of their sSFR does not fully overlaps with the definition of UVJ-passive galaxies usually adopted in literature \citep{2011ApJ...735...86W}, although relaxing the cut in sSFR would improve the agreement, especially for the higher redshift sample. In fact, increasing the cut from log(sSFR [yr$^{-1}$])=$-$9.7 to log(sSFR [yr$^{-1}$])=$-$9.5 (shown as horizontal dashed and dotted dark-grey lines in the bottom panels of Figure~\ref{fig:nf0} respectively), the fraction of commonly selected objects would rise from $\sim55\%$ to $\sim85\%$ for the $z\sim1.23$ sample. Hereafter, we will refer as the "quiescent sample" to the one selected on the basis of the UVJ diagram, unless stated otherwise. Summarizing, the number of galaxies in each sample is 585/285/143/47 and 332/142/83/21 for the comparison/\OII/IR/UVJ subsamples in the F687W17 and F823W17 filters respectively. The definition of the samples is illustrated in Figure~\ref{fig:nf0}. We observe also that our IR detection limit is SFR$\sim$3~M$_{\rm sun}$~yr$^{-1}$ and SFR$\sim$6~M$_{\rm sun}$~yr$^{-1}$, for the F687W17 and F823W17 filters respectively. We mark these constant SFR relation as black-dottet lines in Figure~\ref{fig:nf0}.

\begin{figure}
  \begin{center}
    \includegraphics[width=0.495\textwidth]{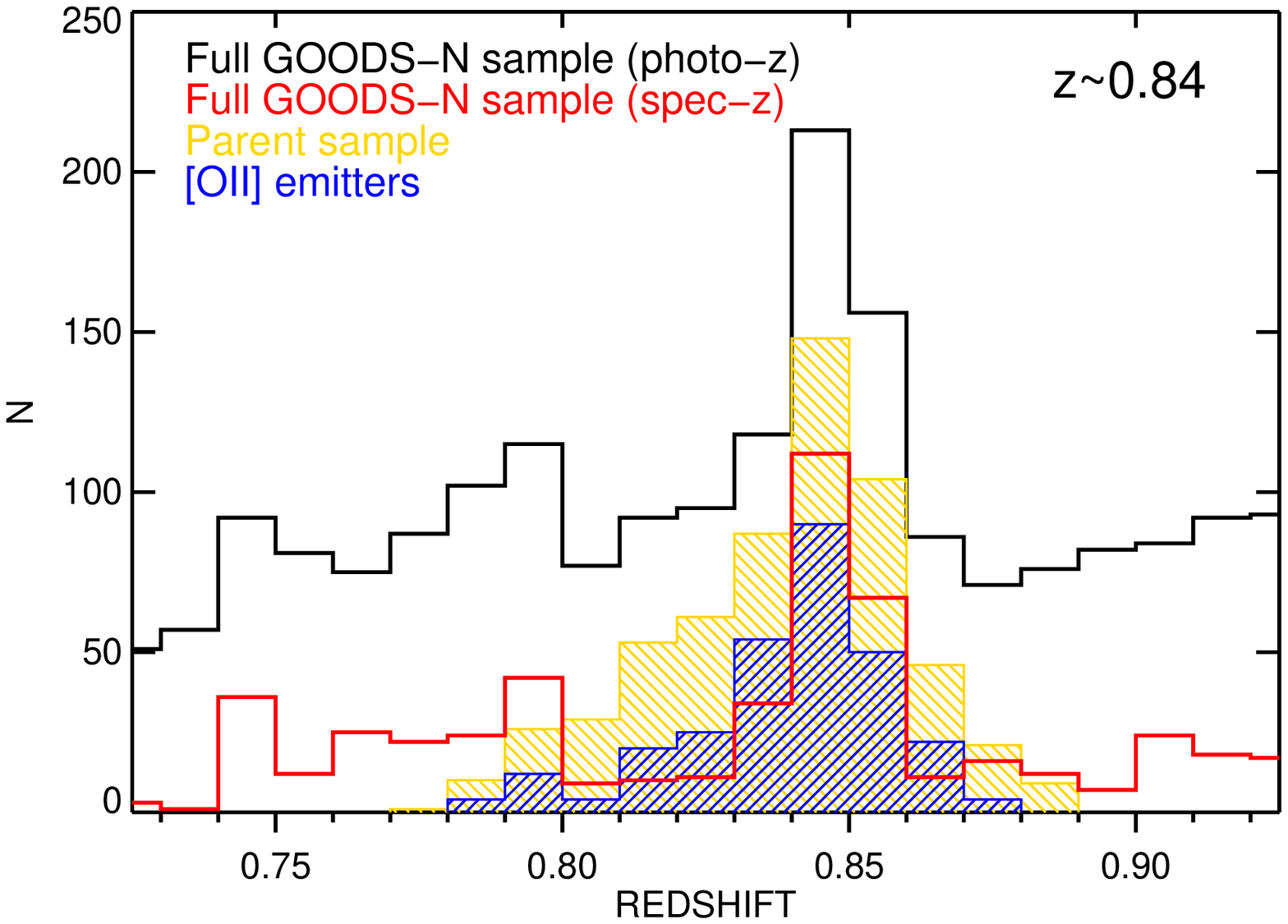}
    \includegraphics[width=0.495\textwidth]{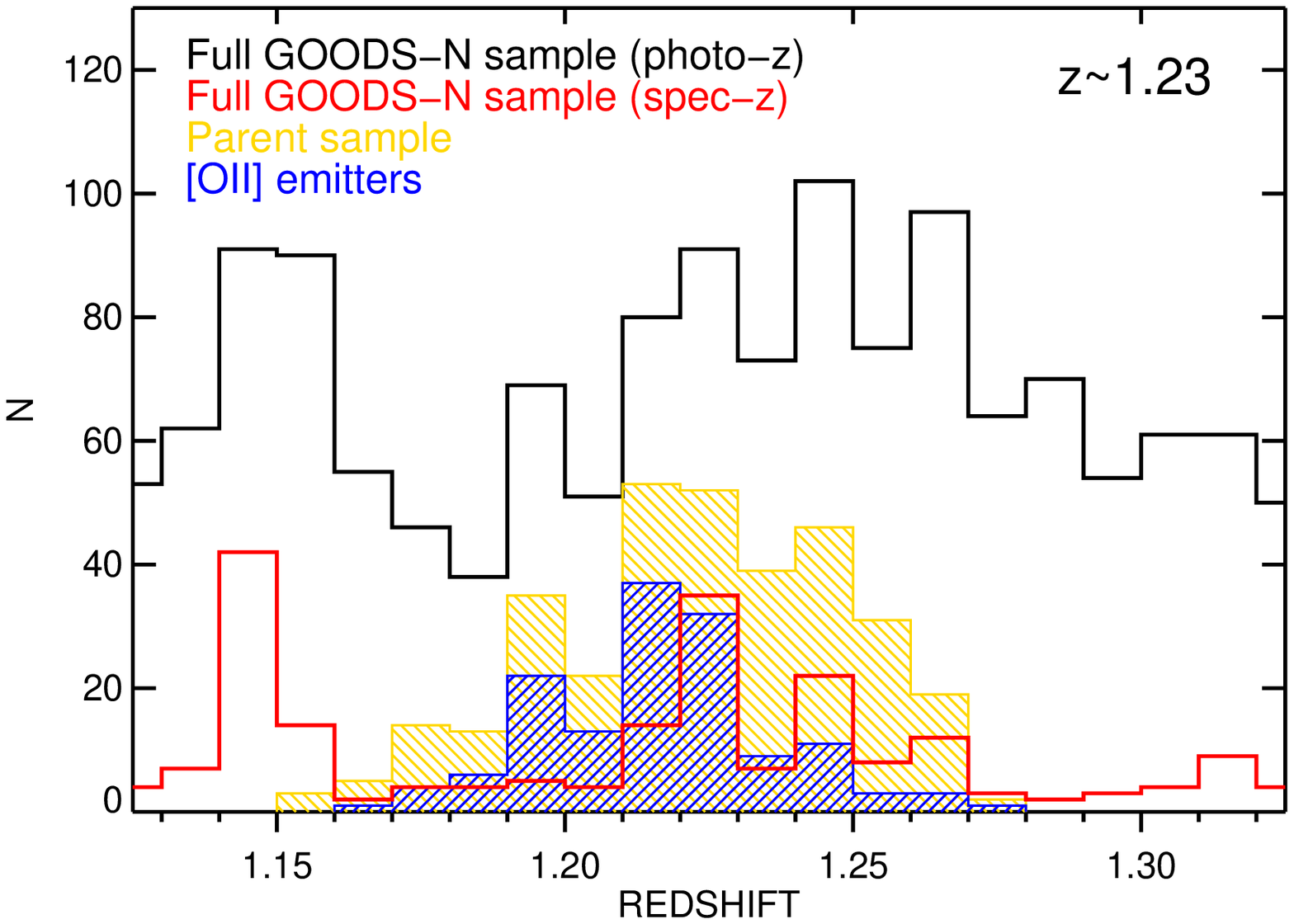}
    \figcaption{\label{fig:check_z} Histograms illustrating the redshift distributions for the 
      selected sample of the \OII\ emitters (blue shaded histogram) and
      the parent sample (yellow shaded histogram) extracted from the 
      parent catalog applying the same filter selection function. 
      For completeness also the photometric and spectroscopic redshift 
      distributions for the parent catalog are shown  
      (black/red outlined histograms respectively). Note
      that for the lower redshift selection (upper panel) the presence of a prominent 
      large scale structure is evident.}
  \end{center}
\end{figure}

In Fig.~\ref{fig:check_z}, we show the spectroscopic and photometric redshifts distributions for the total population of galaxies in GOODS-N extracted from the Rainbow Database. We compare this global distribution with the redshift distribution of {\it bona fide} \OII\ 
emitters identified through the two SHARDS filters used in this work (see Section~\ref{sect:oiisel}) and with the parent sample defined above. From this figure we notice that in the case of filter F687W17, corresponding to the redshift slice at $z\sim0.84$, we are selecting our \OII\ emitters from a large scale overdensed region, clearly identified by the global spectroscopic (and also photometric) redshift distribution. Note also that the photometric and spectroscopic redshift distributions closely agree, confirming the goodness of photometric redshift estimates.

\subsection{AGNs fraction}
\label{sect:AGN}
Our method to detect ELGs is sensitive to both SFGs and AGNs. Since we are interested in the former, we discuss here the possible contamination from AGNs in our sample. Of course, these AGNs likely have also active SF. 

The SHARDS field fully overlaps with the 2~Ms Chandra Deep Field North (CDFN; \citealt{2003AJ....126..539A}). The X-ray source catalog from \cite{2003AJ....126..539A} contains 267 X-ray sources in the area surveyed by SHARDS. The Chandra images are deep enough to detect the X-rays associated with normal SF (e.g., \citealt{2004A&A...427...35P,2010ApJ...724..559L,2011A&A...535A..93P}). In order to
avoid contamination from non-AGNs X-ray sources, we restrict our AGN identification to sources with X-ray luminosities $L_X> 10^{42}~$erg~s$^{-1}$ from \cite{2008ApJ...689..687B} in either the soft (0.5-2 keV) or hard (2-8 keV) band. Cross-matching our samples of \OII\ emitters with the ultra-deep X-ray CDFN catalog, we identify 4 AGNs in the F687W17 sample at {\it z}\,$\sim$0.84 and 2 AGNs in the F823W17 sample at {\it z}\,$\sim$1.23.  These numbers correspond to a fraction of $\sim$1.2\% of the samples, values in good agreement with the X-ray selected AGN fraction of $1-2\%$ estimated by \cite{2009ApJ...701...86Z} in their analysis of {\it z}\,$\sim$1 \OII\ emitters in the DEEP2 survey. \cite{2013ApJ...769...83C} found a slightly higher fraction of AGNs ($\sim3\%$) from their sample of \OII-selected galaxies in the HETDEX pilot survey \citep{2008ASPC..399..115H,2011ApJS..192....5A}, although they used a fainter limit for identifying AGNs from the parent X-ray catalog. As mentioned above, the correlation between X-ray emission and AGN contribution to the derived SFR does not necessarily imply that the integrated \OII\ emission is dominated by flux from the central engine. However, the SFR estimations based on indicators such as the \OII\ emission could be contaminated by the AGN. Indeed, by examining the few X-ray bright objects in our survey we verified that these candidate AGNs are generally not amongst the strongest \OII\ emitters. The uncertainty of contribution of the AGN to the observed \OII\ emission and the small number of objects involved lead us to conclude that the exclusion of candidate AGNs from the final sample constitute a negligible source of uncertainty in our results.
We note that a small fraction of highly obscured AGN could be missed by selecting only on the basis of the observed X-ray emission, which is able to select mostly unobscured and moderately obscured AGN \citep[see, e.g.,][]{2009ApJ...693..447F,2012AdAst2012E...9F}. However, given the extreme depth of the available X-ray data in the GOODS-N field, and the already modest fraction ($\sim1\%$) of unobscured$+$moderately obscured AGN  detected in our sample, we can assume that the additional number of highly obscured AGN is negligible.

\section{Observational properties of the \OII\ emitter sample}
\label{sect:oii-prop}

In this section, we describe the procedure to measure \OII\ fluxes and EWs from the SHARDS data for the \OII\ sample of galaxies described in the previous section. The method is based on the estimation of the continuum emission around the emission line by comparing the SHARDS spectro-photometric data with stellar population models, and then using the central filter to measure the line flux.
This modeling is described in Section~\ref{sect:rainbow}. Then we compare the properties of our \OII\ emitters with other similar samples from the literature.

\subsection{Line fluxes and equivalent widths of \OII\ emitters}
\label{sect:linflux}
For the final samples of \OII\ emitters at {\it z}\,$\sim$0.84 and {\it z}\,$\sim$1.23 selected from the F687W17 and F823W17 filters, we have measured  line fluxes and EW using the SHARDS medium-band data as detailed next. 

The flux emitted in the \OII\ line is by definition (see, e.g., \citealt{2008ApJ...677..169V,2012MNRAS.420.1926S}) :
\begin{equation}
  F(\OII) = (f_\lambda^{central}-f_\lambda^{continuum})\times\Delta\lambda^{central}
\end{equation}
where $f_\lambda^{central}$ and $f_\lambda^{continuum}$ are the measured flux densities in the central filter containing the emission
line and the continuum obtained from the stellar population fits (defined as described below in this section). $\Delta\lambda^{central}$ is the FWHM of the SHARDS central filter. The derived observed equivalent width is:
\begin{equation}
EW(\OII) = F(\OII)/f_\lambda^{continuum}
\end{equation}
where $F(\OII)$ is the line flux defined above.

We have followed a careful and sophisticated method to measure EWs and fluxes, based on a robust determination of the continuum around the emission line. The continuum estimation can be extracted from a linear interpolation between the two continuum filters used in the selection of ELG candidates, but this method can be affected by line contamination of the side filters or other local features. A more accurate estimation relies on the use of stellar population models fitting the medium-band data around the emission line \cite[see, e.g.,][for a similar approach]{2015arXiv150507555D,2015A&A...580A..47V}, thus exploiting all the information available from the SHARDS data.  

\begin{figure}[tb]
  \begin{center}
    \includegraphics[width=0.49\textwidth]{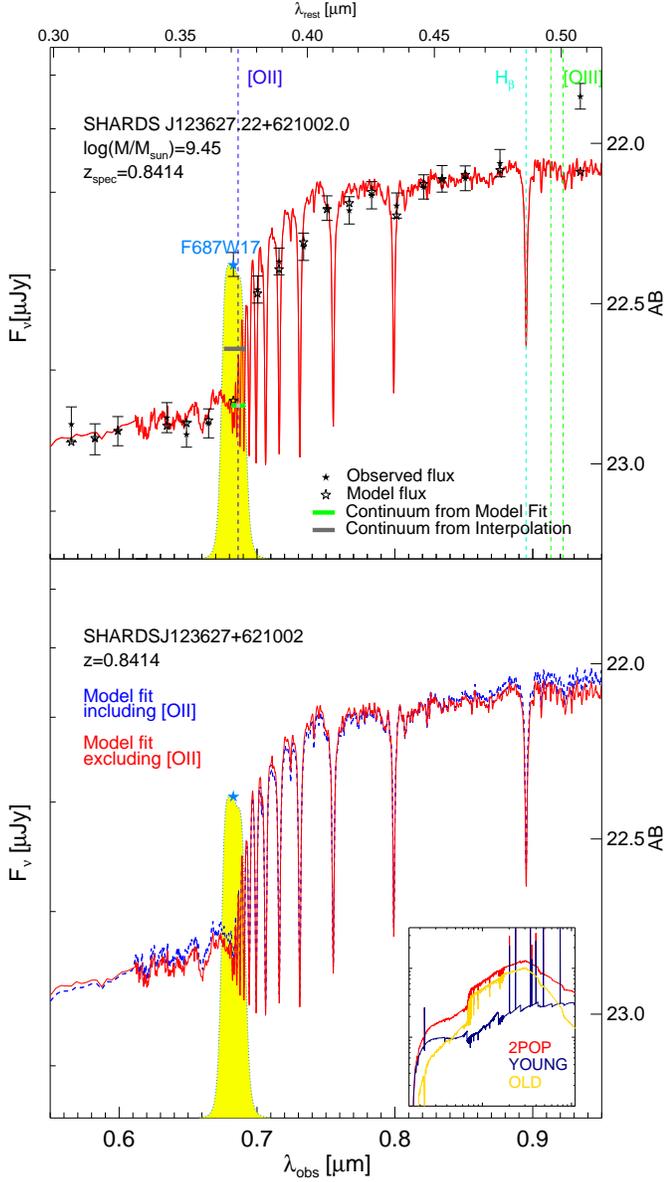}
    \figcaption{\label{fig:multifit} Illustration of the method used
      to refine flux and EW measurements from SHARDS data by
      determining an accurate continuum level ({\it upper panel}) for
      \OII. For all the galaxies selected as \OII\ emitters, we fit the
      SHARDS medium-band data to stellar population models and extract
      the continuum level from the best fitting model. Black filled stars are
      the SHARDS medium-band observed fluxes (with the associated 1$-\sigma$ error bars), 
      with the larger blue star
      indicating the filter used for the detection of the \OII\ emission
      line. In red, we show the best-fit model to the SHARDS data (see
      text for details). Open stars represent the flux predicted from
      the best-fit model, after convolution with the SHARDS filters. 
      We mark with a horizontal grey segment the
      continuum obtained by simple interpolation of the two adjacent
      filters to the central passband. The continuum
      obtained from the model best-fit is marked in green. 
      We also illustrate ({\it lower panel}) the effect of excluding 
      (continuous red line, same as in the upper panel) or including 
      (dashed blue line) the photometric data-point corresponding to the 
      filter containing the \OII\ line. The small inset shows the global two-population 
      fit (in red) depicted in the two main panels, with the
      young/old components overplotted in blue/yellow respectively.
      }
  \end{center}
\end{figure}

In Figure~\ref{fig:multifit}, we illustrate the adopted procedure for a galaxy at {\it z}\,$\sim$0.84. The SHARDS data are fitted to stellar population synthesis models as described in Section~\ref{sect:rainbow}. When performing the fit, we excluded the filter containing the emission line, which is obviously contaminated. The model allows us to determine the continuum at the position of the emission line. The continuum obtained from the best fit (green segment in Figure~\ref{fig:multifit}) is usually lower  (on average $\sim$10-15\% lower density flux) than that obtained from the linear interpolation (grey line) of the two adjacent filters. This translates into systematically
larger (but more robust) EW for the line (on average $\sim$35-40\% larger EWs) and higher fluxes (on average $\sim$25-30\% larger fluxes). In the lower panel of Figure~\ref{fig:multifit}, we show the effect of including or excluding the filter that contains the emission line.  Typically, there is no much difference between the two fits ($<10\%$ difference in derived EW), but larger discrepancies (up to $\sim50\%$ difference in derived EW) can be found when the emission line is very strong, affecting the continuum determination.

\begin{figure}[tb]
  \begin{center}
   \includegraphics[width=1.\columnwidth]{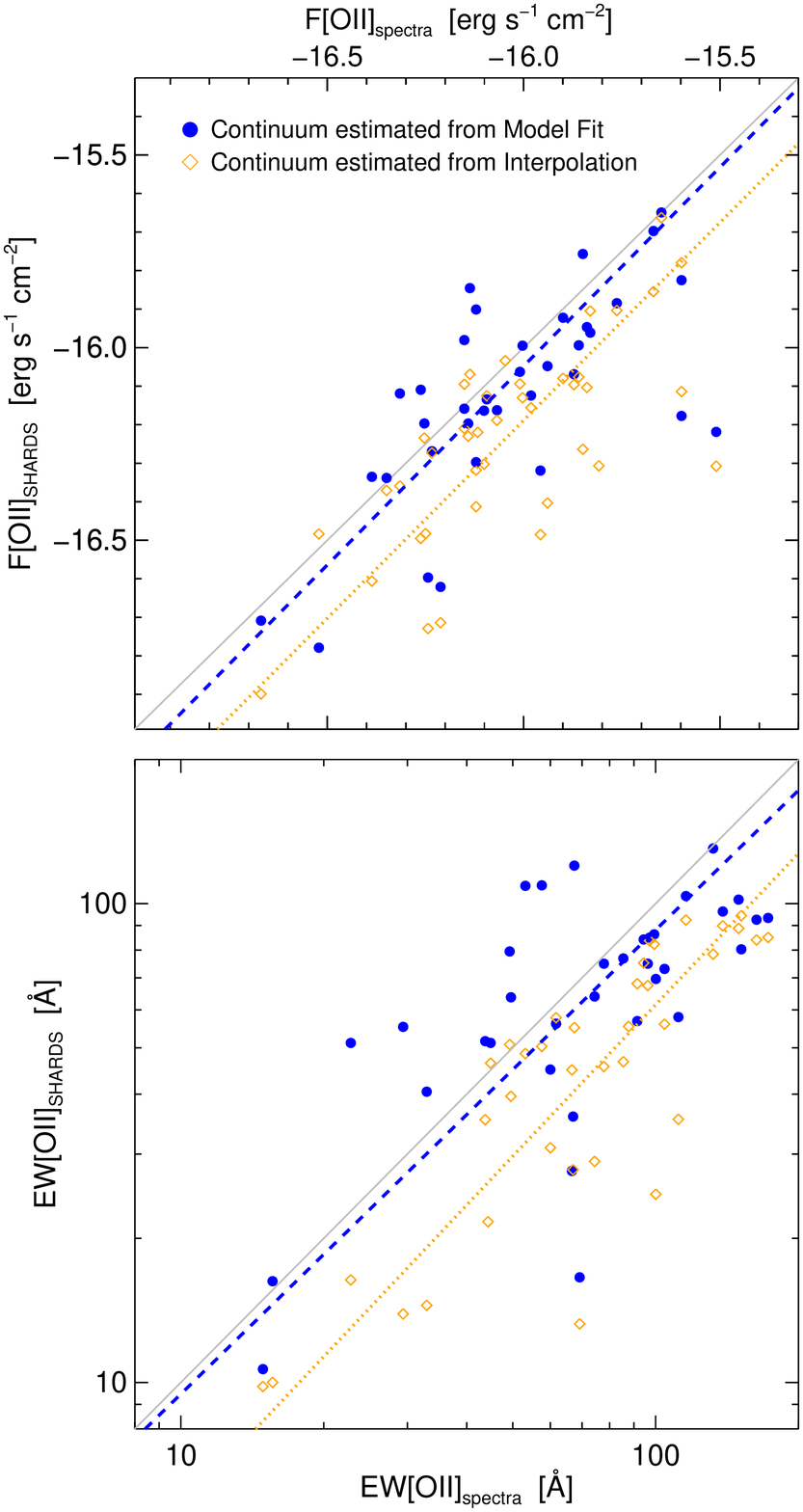}
   \figcaption{\label{fig:fluxcheck} Comparison between spectroscopic- and
     SHARDS-derived line fluxes and EWs for the \OII\ line (upper and lower panel respectively).
     Spectroscopic measurements are obtained from public data (TKRS and DEEP3
     spectra). The continuum grey line represents the identity relation. 
     Blue points and orange diamonds represent flux and EW estimates based on best-fit or 
     linear interpolation for the continuum, respectively. 
     The blue-dashed and orange-dotted lines are the linear best-fit to the two datasets. }
  \end{center}
\end{figure}

In order to assess the reliability of our measurements based on spectro-photometric data from SHARDS, we compare line fluxes derived
from our medium-band filters to those derived from spectra of the same objects and available from literature (TKRS, \citealt{2004AJ....127.3121W}, and DEEP3 data, \citealt{2011ApJS..193...14C}). The subsample of galaxies having spectra of good quality, i.e. $S/N$ high
enough to robustly measure EW(\OII) and F(\OII), contains 42 galaxies ($\sim$10\% of the final sample). We show the comparison between the line fluxes measured on the basis of the publicly available spectroscopic and SHARDS photometric data in Figure~\ref{fig:fluxcheck}. 

On average, fluxes measured from spectroscopy are $\sim$30-35\% larger than fluxes measured from SHARDS photometry when using a continuum determined from linear interpolation between adjacent filters. Our procedure to estimate the continuum based on stellar population synthesis models significantly improves the comparison between spectroscopic and photometric line fluxes, making them consistent within $\sim$5-10\%. Despite the large scatter,  this result supports the robustness of our approach.  Note that the spectroscopic measurement of the line flux is affected by slit losses, and typically this correction is carried out by assuming that the whole galaxy has a constant EW equal to the one measured in the spectra. If galaxies count with nuclear bursts, this correction would overestimate the total flux of the galaxy. This is confirmed by Figure 9, where we show that the observed EW(\OII) from SHARDS data are, on average, $\sim$30\% smaller than those measured in spectra. In contrast, photometric measurements of the line flux using integrated magnitudes are not affected by aperture effects, although spatial gradients of the emission could affect the selection of ELG candidates.

\begin{figure*}[tb]
  \begin{center}
   \includegraphics[width=1.\textwidth]{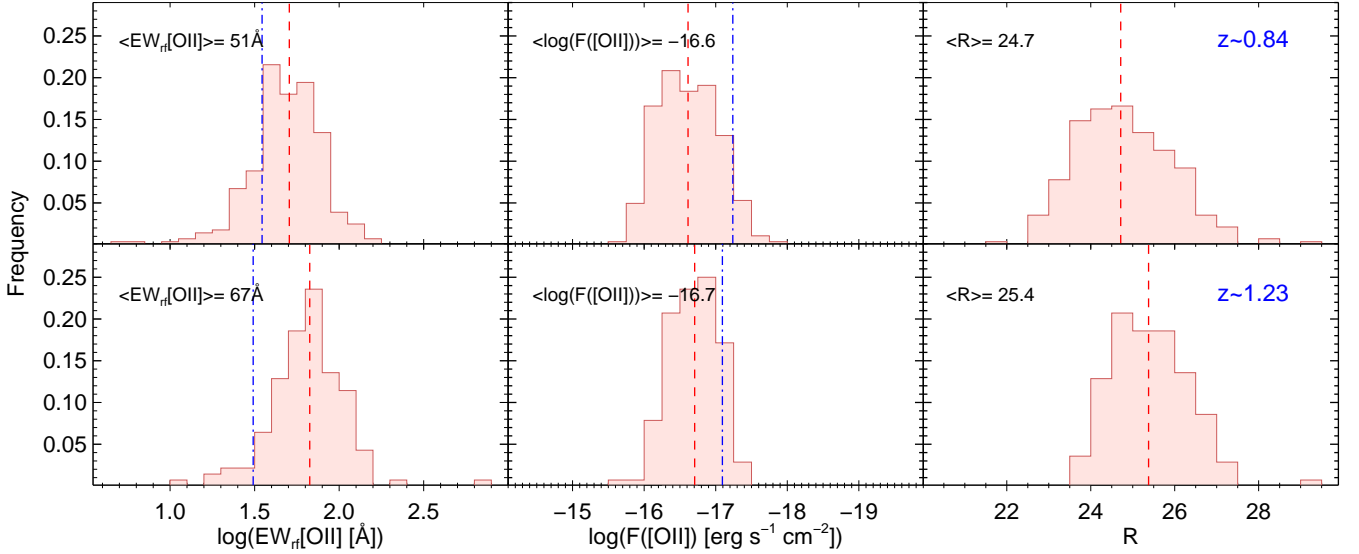}
   \figcaption{\label{fig:multi_EW} Distributions of EWs, line fluxes and
     R-band magnitude for the \OII-selected galaxies at {\it z}\,$\sim$0.84 (upper
     panels) and {\it z}\,$\sim$1.23 (bottom panels). The vertical dashed lines represent the median values also given in each panel. The blue dot-dashed line in the left and central panels marks the minimum \EW\ and corresponding flux achievable assuming the typical uncertainty for the faintest galaxies in our sample.}
  \end{center}
\end{figure*}

We show the resulting distribution of rest-frame EWs (\EW), F(\OII) and $R$-band magnitudes for the final sample in Figure~\ref{fig:multi_EW}. Errors in the estimates of the line flux and continuum level propagate into uncertainties in the derived \EW\ of $\sim$20\% at most. We observe a median  (and [quartiles]) rest-frame \EW=51[38,65]\,\AA\ at {\it z}\,$\sim$0.84 and \EW=67[49,91]\,\AA\ at
{\it z}\,$\sim$1.23. A $\sim$0.5~mag shift is also seen in the observed $R$-band magnitude distribution, with the higher redshift sample peaking at fainter magnitudes. The line flux distribution is similar in the two cases, ensuring that we are not introducing any particular
redshift-dependent bias in line flux measurements. The median observed line flux for both distributions is $\sim$2$\times$$10^{-17}$~erg$/$s$/$cm$^2$, with measurements extending to $\sim$2$\times$$10^{-18}$~erg$/$s$/$cm$^2$ on the faint
end. We note that the minimum rest-frame equivalent width achievable assuming the typical uncertainty for the faintest \OII-selected galaxies corresponds to 30\,\AA\ and 36\,\AA\ for the higher and lower redshift sample respectively. For brighter galaxies with smaller uncertainties we can reach even smaller values for the EW. The smaller value for the higher redshift sample is due to the smaller rest-frame filter width at this redshift. These values are indicated as a dot-dashed vertical blue line in the left panels of Figure\,\ref{fig:multi_EW}. These lower limits ensure a robust determination of the median \EW\ values discussed in the following section. Indeed, we can convert these values to a lower limit in the line-flux detection (F(\OII)$_{\mathrm{limit}}$$\sim$6--8$\times$10$^{-18}$\,erg\,s$^{-1}$\,cm$^{-2}$), also reported as vertical blue dot-dashed line in the central panels of Figure\,\ref{fig:multi_EW}. In this case, as expected, a slightly higher observed line flux limit is found for the higher redshift sample.
\begin{figure}[tb]
  \begin{center}
    \includegraphics[width=.98\columnwidth]{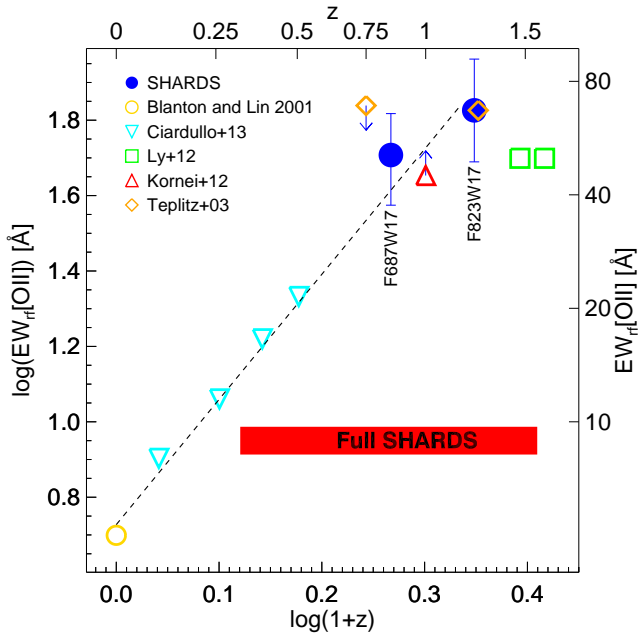}
  \end{center}
  \figcaption{\label{fig:EW_evol} Evolution of the median \EW\ with redshift. 
    We represent the values derived in this study for {\it z}\,= 0.84 and {\it z}\,= 1.23 and a compilation from
    the literature at different redshifts up to {\it z}\,$\sim$1.5 (see the legend). 
    The dashed line represents the best fit in the form \EW$\sim(1+z)^\alpha$, with $\alpha=3.2$. 
    The arrows indicate upper or lower limits respectively (see text for
    details). The datum at {\it z}\,$\sim$0.75 from \citet{2003ApJ...589..704T} 
    should be considered as an upper limit since their
    detection limit does not allow to measure EWs lower than
    $\sim35$\AA. The point from \citet{2012ApJ...758..135K} should be
    considered as a (slightly) lower limit since their EW distribution
    is biased toward higher mass objects
    ($<$M$>\sim$1.5$\times$10$^{10}$M$_{\mathrm{sun}}$). The data
    from \citet{2012ApJ...757...63L} show a lower value ($\sim50$\AA) of the
    median EW measured for their two redshift bins at {\it z}\,$\sim$1.47 and
    {\it z}\,$\sim$1.62 with respect to the trend presented in the lower
    redshift bins.  These last points are not included in the fit, because they 
    may represent a flattening of the \EW\ evolution at $z>1$.}
\end{figure}

These flux and EW detection limits are comparable to those obtained with deep spectroscopy. We conclude that our SHARDS spectro-photometric survey is very effective in selecting ELGs down to the faintest flux levels achieved by spectroscopy. Note that the median fluxes and EWs of our samples are significantly higher than the detection limits, so the results presented in the following sections are not significantly affected by incompleteness at the faintest flux or smallest EW levels.

\subsection{{\rm{\EW}} evolution}
\label{sect:ew-evol}
The measurement of the EW(\OII) provides a valuable source of information on the relevance of the ongoing star formation in galaxies. As it is well known (e.g., Kennicutt 1998) the \OII\ line flux is strictly related to the radiation field coming from the young ($t<20$~Myr) and massive stellar (M$>8$~M$_{\mathrm{sun}}$) populations (and also to metallicity and other properties such as density and temperature). On the other hand, the continuum is more representative of a larger timescale (t$\sim$1~Gyr) star formation. The ratio of these quantities, i.e. the EW(\OII), thus provides an estimate of the current SF efficiency at the epoch of observations with respect to the average SF in the past galaxy life. This is an indicator of the current galaxy SF efficiency and it is largely independent on the degree of internal extinction of a galaxy. 

In some recent works \citep[see, e.g.,][]{2013ApJ...769...83C,2012PASP..124..782L,2012ApJ...758..135K,2012ApJ...757...63L} the distribution of the EW for large samples of \OII\ ELGs  are presented and their median values given for samples ranging from {\it z}\,$=$0 to {\it z}\,$\sim$1. Locally, the distribution of rest-frame EW(\OII) for \OII-selected galaxies is known to peak near $5$~\AA, and then it rapidly decays without any noticeable dependence on the galaxy luminosity \citep{2000ApJ...543L.125B}.
\begin{figure}[tb]
  \begin{center}
   \includegraphics[width=1.\columnwidth]{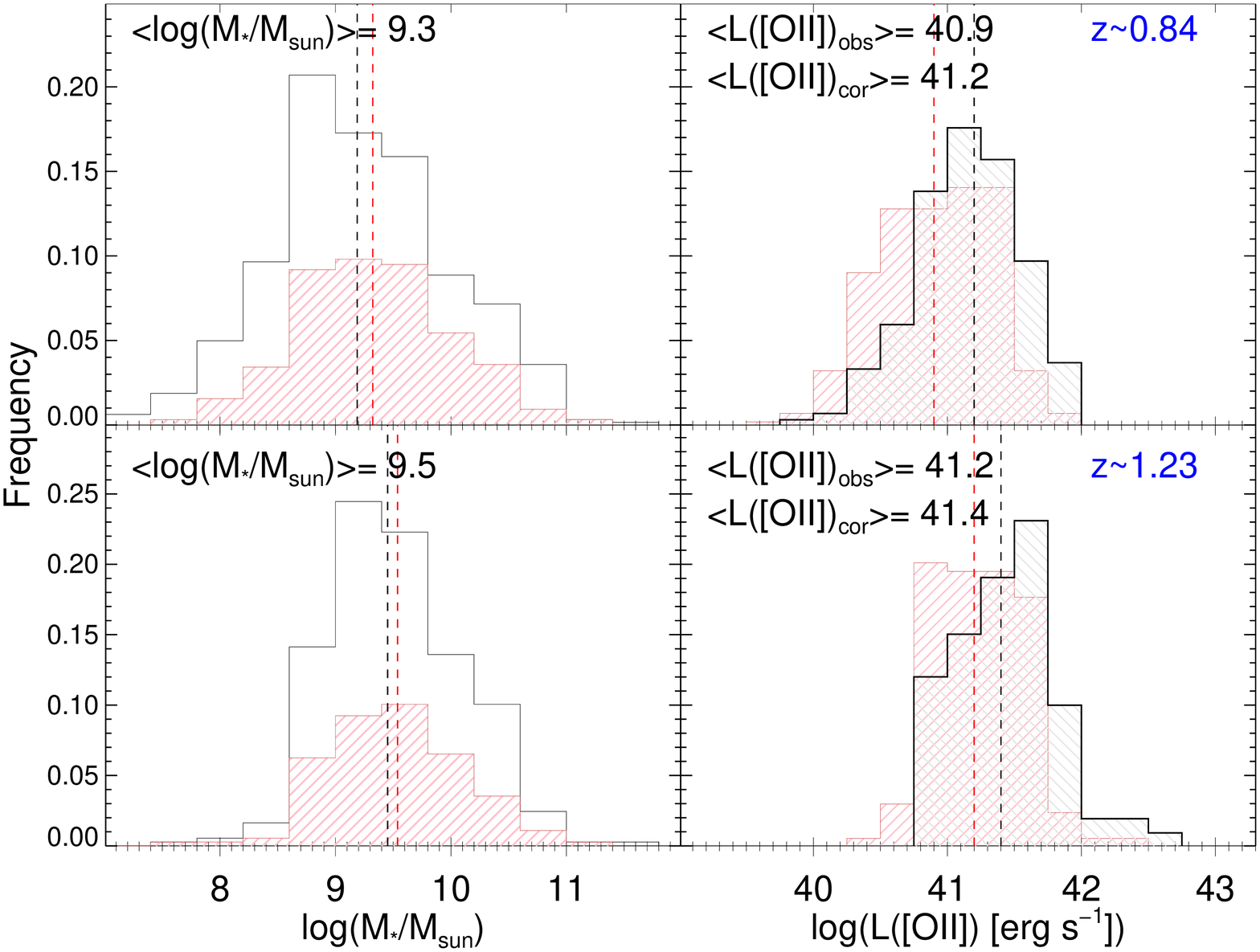}
   \figcaption{\label{fig:multi_prop} {\it Left panels:} distribution of stellar masses
   for the \OII\ emitters (hatched histogram) and the parent sample (black outlined histogram).
   Vertical dashed lines indicate the median value for the \OII\ (red) and the comparison (black) sample respectively.
     {\it Right panels:} observed (hatched rose histogram) and extinction-corrected (hatched grey histogram)
     L(\OII) luminosities. Vertical dashed lines indicate the
     median value for the observed (red) and extinction-corrected (black) luminosity distributions.}
  \end{center}
\end{figure}

In Figure~\ref{fig:EW_evol} we show the evolution of the median \EW\ measurements of \OII-selected samples from literature at redshifts 0$<$\,{\it z}\,$<$1.5 compared to the values obtained for our SHARDS emitters. Our estimates of the \EW\ are consistent with those derived by other authors at similar redshifts (compare them with those from \citealt{2012ApJ...758..135K} and \citealt{2003ApJ...589..704T} in the figure), meaning that our technique based on medium-band data does not include any systematic bias on these
measurements and provides robust EW values.

The \EW\ evolution curve appears to rise as $\sim$(1+{\it z})$^3$ up to {\it z}\,$\sim$1, and then flattens or starts to decrease at higher redshifts.
We interpret this as a real decrement presented by the \EW\ evolution curve. In fact, the global behaviour of the evolution of the \EW\ with redshift appears to follow that of the SFR density (SFRD) of the Universe \cite[see e.g.][for a recent review]{2014ARA&A..52..415M}, presenting a steep rise (proportional to $\sim$(1+{\it z})$^\alpha$, with $\alpha$$\sim$3.2), up to {\it z}\,$\sim$1 and a flattening (or slow decline) at higher redshift (as suggested by the measurements at {\it z}\,$\sim$1.5 from \citet{2012ApJ...757...63L}). Indeed, the value of the slope $\alpha$ is quite similar to those derived form SFRD studies, where values between 3 and 4 are usually obtainded (e.g.
\citealt{1996ApJ...460L...1L,2003ApJ...593..258B,2005ApJ...630...82P,2007A&A...472..403T,2002MNRAS.337..369T,2012A&A...539A..31C}). This trend reflects the varying SF efficiency at different epochs in the life of the Universe, which is well traced by such a simple quantity as the EW of \OII\ emitters.  

\section{Physical properties of the \OII\ emitters at \MakeLowercase{{\it z}}\,$\sim$0.84 and \MakeLowercase{{\it z}}$\,\sim$1.23}
\label{sect:oii-sfr}
In this Section, we discuss and compare the SFR, stellar mass and (mass-weighted) age properties of the \OII\ ELGs selected in the redshift bins identified by our two SHARDS filters. In the following Section we will compare the properties of \OII\ emitters with the other subsamples defined in Section~\ref{sect:complem}, investigating the effect of using different SFG selection techniques.

\begin{figure}[tb]
  \begin{center}
   \includegraphics[width=1.\columnwidth]{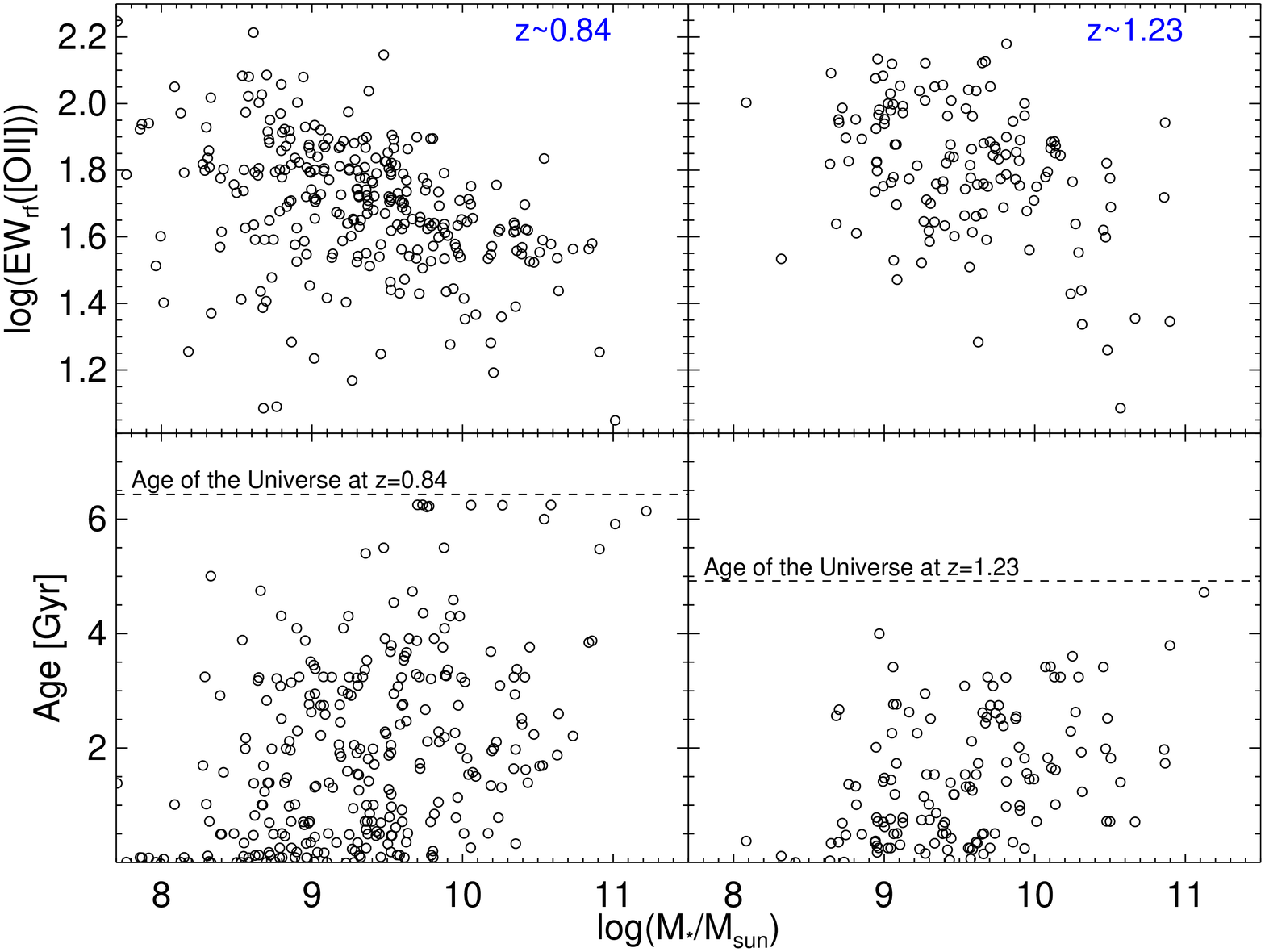}
   \figcaption{\label{fig:multi_age} Relations between (mass-weighted) stellar ages, stellar masses, and \EW\ for the ${\it z} = 0.84$ (left panels) and ${\it z} = 1.23$ \OII\ emitters. These quantities were derived from two-populations SED fitting models that assumed a Chabrier IMF, exponentially declining SFHs, and \cite{2000ApJ...533..682C} internal dust reddening. See text for details on the SED modeling. 
       }
  \end{center}
\end{figure}

The stellar mass and the \OII\ observed (not corrected for dust extinction) luminosity distributions, L(\OII)$_{\mathrm{\mathrm{obs}}}$, for the two samples of SHARDS \OII\ emitters at redshift $\sim0.84$ and $\sim1.23$ are shown in Figure~\ref{fig:multi_prop} (left panels). The average values of the stellar mass in the two redshift bins are similar, passing from $\langle$log(M/M$_{\mathrm{sun}}$)$\rangle\sim9.3$ at {\it z}\,$\sim$0.84 to $\langle$log(M/M$_{\mathrm{sun}}$)$\rangle\sim9.5$ at {\it z}\,$\sim$1.23 (left panels). We also note that the average stellar mass is very similar to that obtained for the parent samples for the two redshift intervals (overplotted as black histogram). 
\begin{figure*}[tb]
  \begin{center}
    \includegraphics[width=1.\textwidth]{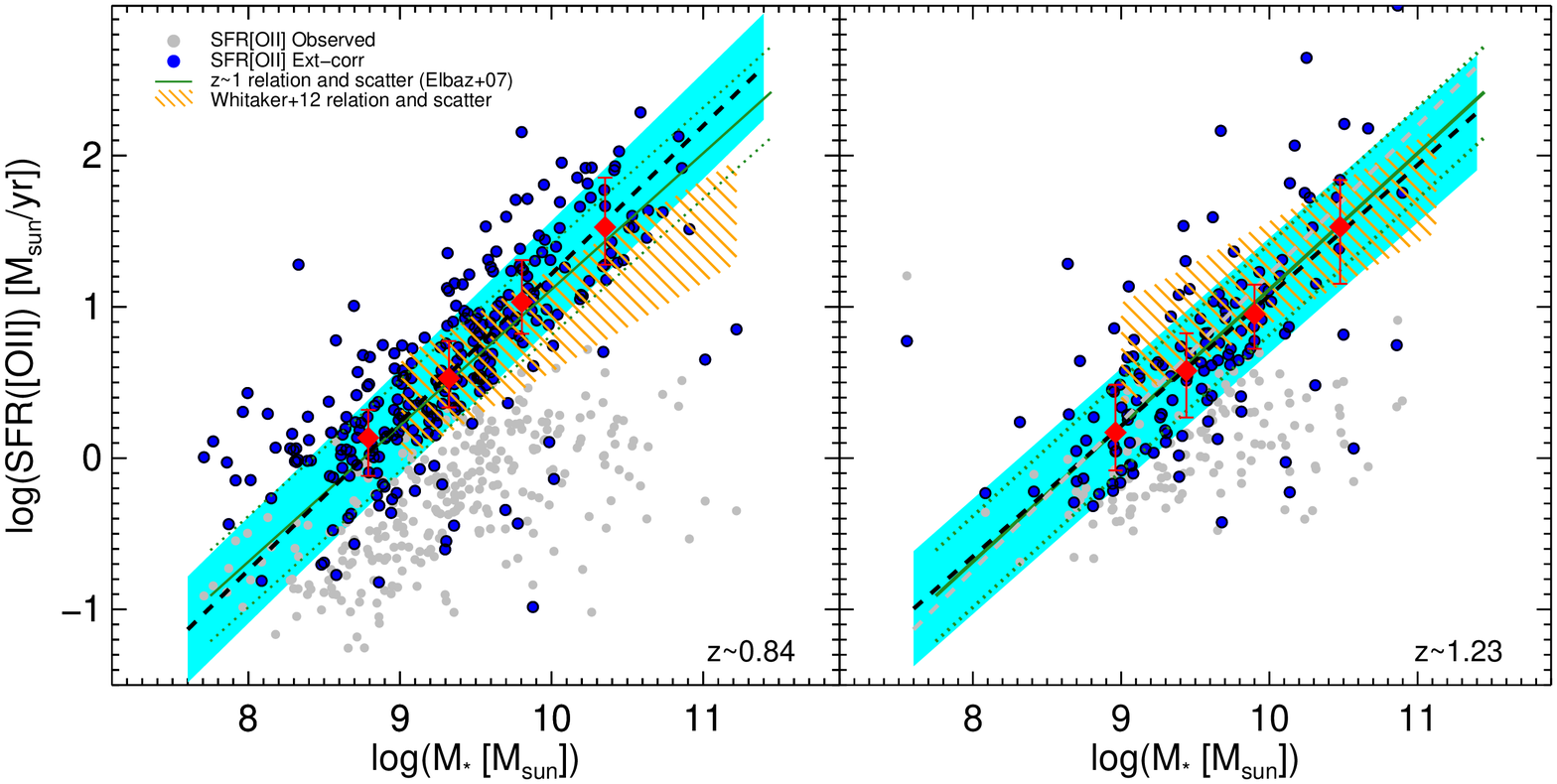}
    \figcaption{\label{fig:SFRcomp_F687} 
      SFR derived from the \OII\ line versus stellar mass relation at redshift {\it z}\,$\sim$0.84
      (left panel) and {\it z}\,$\sim$1.23 (right panel). Observed SFR(\OII) 
      (grey filled dots) and extinction-corrected SFR(\OII) (blue open dots) 
      are shown in the plot. The black dashed line is the fit to our data, 
      with the scatter in the relation shown as cyan shaded area. 
      For comparison, we also overplot, as red diamonds with error bars, the median 
      and (first and third) quartiles for the four equally populated 
      mass bins including 90\% of data (i.e. clipping extremes).
      The orange shaded area indicate the
      SFR-M$_*$ relation (and its scatter) from Whitaker et al. (2012) for the two redshift bins. 
      The green continuum and dotted lines represent the SFR-M$_*$ at redshift z$\sim1$ and its 
      68\% confidence interval, respectively, from Elbaz et al. 2007. 
      In the right panel, we also depict as a grey dashed line the SFR-M$_*$ relation from the left panel.
      }
  \end{center}
\end{figure*}

In the right panels of Fig.~\ref{fig:multi_prop} we show instead the distribution of the observed and extinction-corrected \OII\ luminosities (L(\OII)$_{\mathrm{obs}}$ and L(\OII)$_{\mathrm{cor}}$), adopting the extinction values derived from our two-population SED-fitting technique. The average optical extinction values, for the old and young populations, are A$_{old}\sim$0.9 and A$_{you}\sim$1.5, respectively. We adopt the extinction from the young population to correct the \OII\ luminosities, since \OII\ emission is expected to be mostly produced in young star-forming regions. A \cite{2000ApJ...533..682C} attenuation law is assumed, with a factor $\sim$2.3 accounting for the conversion from the stellar to the gas extinction. This conversion factor has been derived for a sample local star-forming galaxies, while recent studies have suggested that it might be evolving with redshift \cite[see, e.,g.,][]{2011ApJ...742...96W,2013ApJ...777L...8K,2015ApJ...807..141P,2015arXiv150801679T}. Nonetheless, due to the large uncertainties in the derivation of this factor we preferred to use the classical value for local galaxies.

The median values for L(\OII)$_{\mathrm{obs}}$ are $\sim$$10^{40.9}$~erg~s$^{-1}$ and $\sim$10$^{41.2}$~erg~s$^{-1}$ for the F687W17 and F823W17 samples respectively, that shift to $\sim$$10^{41.2}$~erg~s$^{-1}$ and $\sim$10$^{41.4}$~erg~s$^{-1}$ for  L(\OII)$_{\mathrm{cor}}$. The median values for the observed and corrected \OII\ luminosity distributions increase about $\sim$0.2--0.3~dex from the lower to the higher redshift bin.

In Figure~\ref{fig:multi_age} we plot the M$_*$-age and M$_*$-\EW\ diagrams for the two samples of \OII-selected ELGs at {\it z}\,$\sim$0.84 and $\sim$1.23.  Despite the large scatter and intrinsic uncertainties present in the diagrams, we can draw some conclusion about the relation of \OII\ equivalent widths, stellar populations ages and galaxy stellar masses. More specifically, if we divide our samples in subsamples containing galaxies above or below the median value of the \EW, we see that the median value of the mass is $\sim$4 ($\sim$2) times larger for galaxies in the low-\EW\ sample with respect to galaxies belonging to the high-\EW\ sample at {\it z}\,$\sim$0.84 ({\it z}\,$\sim$1.23). 

Regarding the ages of the stellar populations we do not see any specific trend related to the EW of the \OII-selected galaxies. Instead, we
see that galaxies with masses below the median mass value have on average younger ages, by a factor of $\sim$2 ($\sim$4), with respect to the more massive galaxies for the {\it z}\,$\sim$0.84 ({\it z}\,$\sim$1.23) sample, as might be expected within the {\it downsizing} scenario of galaxy evolution, where massive galaxies form the bulk of their mass at an early stage of evolution. These results are in qualitative agreement e.g. with those from \citet{2012ApJ...757...63L}, although we find larger average stellar ages perhaps due to our fitting method including two populations (thus being less affected by recent bursts of SF).

\begin{figure*}[tb]
  \begin{center}
\includegraphics[width=1.\textwidth]{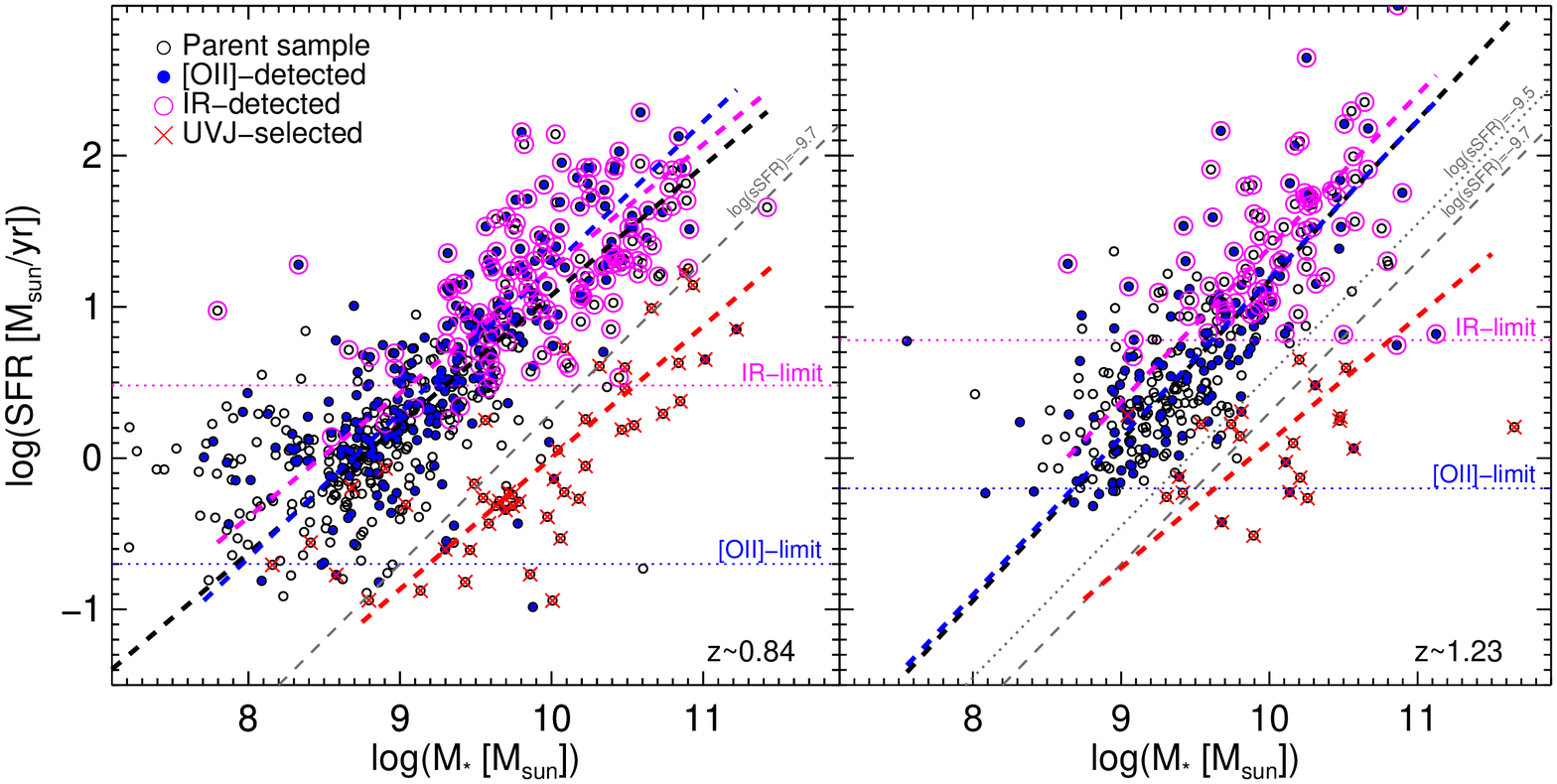}
    \figcaption{\label{fig:nf1} Star formation rate versus stellar mass diagrams (as derived from SED fitting) for the different sample selections indicated in the legend. The dashed lines, identified by the color corresponding to the symbols in the legend, represent the linear fits to the SFR-M$_*$ relation for each subsample. The dark-grey dashed and dotted lines give the the SFR-M$_*$  for galaxies with a constant log(sSFR [yr$^{-1}$])=$-$9.7 or $-$9.5 respectively. See the text for details.}
  \end{center}
\end{figure*}

Finally, in Figure~\ref{fig:SFRcomp_F687}, we plot the SFR-M$_*$ diagram for our \OII\ samples. In these plots we correct (assuming for the dust attenuation the prescription detailed below) the SFR derived from the observed L([OII])$_{\mathrm{obs}}$ through the \cite{2004AJ....127.2002K} relation: 
\begin{equation}
\begin{split}
\mathrm{SFR_{obs}}&\mathrm{(\OII)=}\\
& \mathrm{(1.4\pm0.4)\times10^{-41} L(\OII)~[M_{sun}~yr^{-1}]}
\end{split}
\end{equation}
and scale the SFR from Salpeter to Chabrier IMF by dividing by 1.7. 

We derive the extinction from the ratio of total SFR to the observed SFR$_{\mathrm{obs}}$(\OII), where the total SFR is a combination
of the obscured and unobscured contributions represented by the UV and IR SFR estimates (or predicted values from best SED-fits when dealing with IR non-detections). 
We adopt here the parametrisation introduced by Iglesias-Par{\'a}mo et al. (2006, see also \citealt{2003ApJ...586..794B,2003A&A...410...83H,2004A&A...419..109I,2011ApJ...727...83M}):
\begin{equation}
\mathrm{SFR=SFR_{\mathrm{obs}}(UV)+(1-\eta)\times SFR_{\mathrm{obs}}(IR)}
\end{equation} 
\noindent which includes a correction factor $(1-\eta)$ accounting for the non-negligible contribution to dust heating coming from old stellar populations and that would bias the SFR estimate towards higher values. We assume $\eta=0.32$ and $\eta=0.09$ 
depending on the IR luminosity being lower than $10^{11}$L$_{\mathrm{sun}}$ and higher than $10^{11}$L$_{\mathrm{sun}}$ 
respectively \cite[see][]{2011ApJ...727...83M}.

The observed SFRs were estimated for all galaxies in the parent sample based on either mid- and far-IR data from Spitzer and Herschel, or from UV luminosities. GOODS-N has been observed with the deepest MIPS data in the sky, with a 5$\sigma$ limit of 30$\mu$Jy \citep{2005ApJ...630...82P}, and also very deep PACS and SPIRE observations, with 5$\sigma$ limits of 1.7~mJy and 9~mJy at 100$\mu$m and 250$\mu$m, respectively. These limits (the deepest being MIPS 24$\mu$m data) correspond roughly to 5-10 M$_{\rm sun}$~yr$^{-1}$ for galaxies at 1.0$<$z$<$1.5 \citep{2005ApJ...630...82P,2011A&A...533A.119E}.

The total infrared luminosity, L(IR[8-1000]), has been estimated from SED-fitting of the MIR-to-FIR photometry using the \cite{2001ApJ...556..562C}, \cite{2002ApJ...576..159D}, and \cite{2009ApJ...692..556R} templates. The infrared luminosities are then converted into SFRs using the \cite{1998ARA&A..36..189K} relation :
\begin{equation}
{\rm SFR_{\mathrm{obs}}(IR) =2.7\times10^{-44} L(IR)~[M_{sun}~yr^{-1}]}
\end{equation}
normalized to a Chabrier IMF. 
In order to estimate SFRs for galaxies not detected in the mid-or far-IR (including the quiescent galaxy sample), we used UV luminosities at rest-frame 280~nm. These luminosities were converted into SFRs by applying \cite{1998ARA&A..36..189K} equation normalized to a Chabrier IMF:
\begin{equation}
{\rm SFR_{obs}(UV)=0.95\times 10^{-28} L(UV)~[M_{sun}~yr^{-1}] }
\end{equation}
The UV-based SFRs were corrected for attenuation following the recipe from \cite{1999ApJ...521...64M}, based on the UV/IR-$\beta$ relation (IRX-$\beta$). The UV slope $\beta$ for each galaxy was measured from the SEDs, interpolating the best fitting models between 150 and 300~nm.

We compare our results with the SFR-M$_*$ relation from \cite{2012ApJ...754L..29W}, that studied the evolution of the SFR-M$_*$ relation up to redshift 2.5 for a sample of $>20000$ galaxies from NEWFIRM Medium-Band Survey, shown as the orange hatched region.  
Comparing the linear fit to our data (dashed black lines in Figure~\ref{fig:SFRcomp_F687}) to the SFR-M$_*$ relation at the same redshift from \cite{2012ApJ...754L..29W} we observe a slightly larger slope in our case. This can be due to the different selections applied in the two cases, in fact a larger slope is also found by \cite{2012ApJ...754L..29W} when selecting only blue SF galaxies, which might be more comparable with our \OII\ selection. 

Indeed, we should mention that there could be metallicity effects, in the sense that for higher masses we expect higher metallicities, and then we can have larger L(\OII) for the same SFR. In this sense, our SFR(\OII) could be overestimated at large masses. 

As check, we have also derived SFR(\OII) using the mass-dependent calibration for the dust-extinction from \cite{2010MNRAS.409..421G}. Using this relation, we would recover a better agreement with the results from \cite{2012ApJ...754L..29W}, in terms of slope of the massive end of the SFR-M$_*$ diagram. 

On the other hand, we compare our SFR-M$_*$ relation with the one presented by \cite{2007A&A...468...33E} for redshift z$\sim1$ SF galaxies finding very good agreement between their logarithmic slope of $\sim0.9$ and our best linear fit providing a logarithmic slope of $\sim0.98$ at $z\sim0.84$ and $\sim0.86$ at $z\sim1.23$, and a scatter of $\sim$0.35 and $\sim$0.38, respectively, at these two redshifts. 

As noted in \cite{2012ApJ...754L..29W}  a flattening of the SFR-M$_*$ relation moving toward higher redshifts has been observed in several previous works \citep[see e.g.][]{2007ApJ...660L..43N,2011ApJ...730...61K} and confirmed using our \OII-selected sample of SF galaxies using SHARDS data. Nonetheless, as also commented in Section~\ref{sect:intro}, various effects can contribute to bias these results, leading to a flattening of the SFR-M$_*$ relation, especially for the higher redshift samples, where the mass completeness may represent a major issue.
\begin{figure*}[tb]
  \begin{center}
    \includegraphics[width=0.33\textwidth]{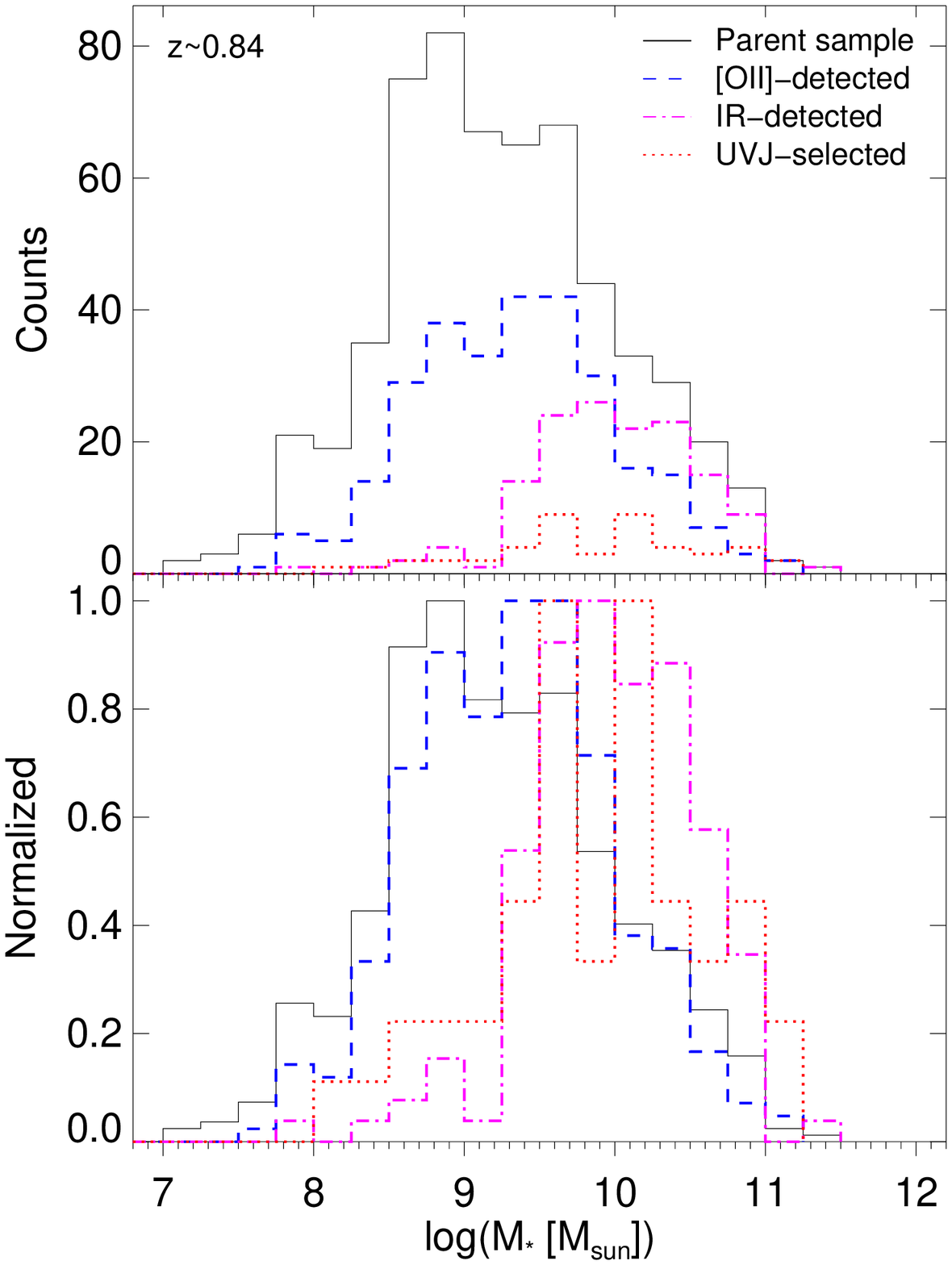}
    \includegraphics[width=0.33\textwidth]{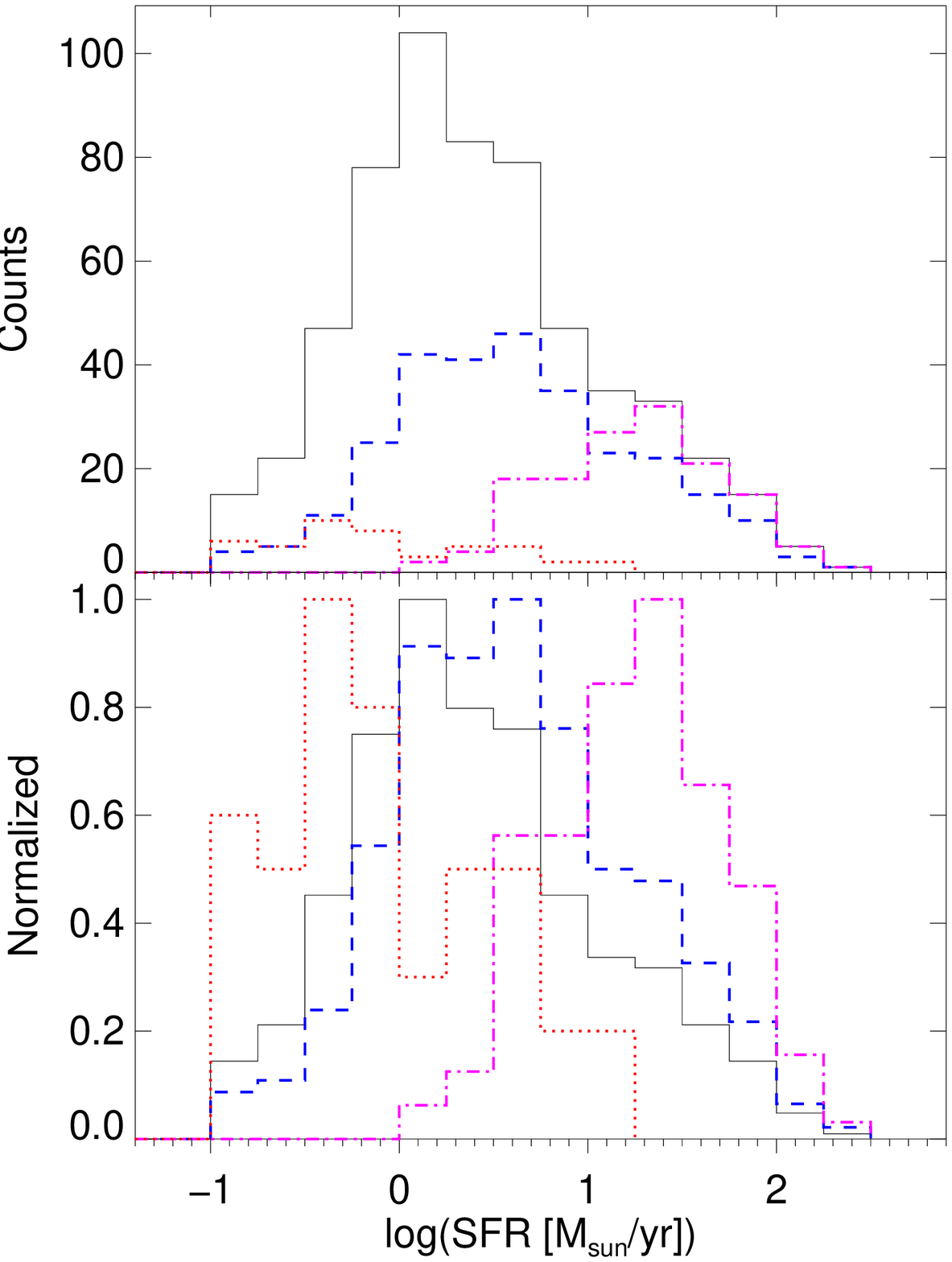}
    \includegraphics[width=0.33\textwidth]{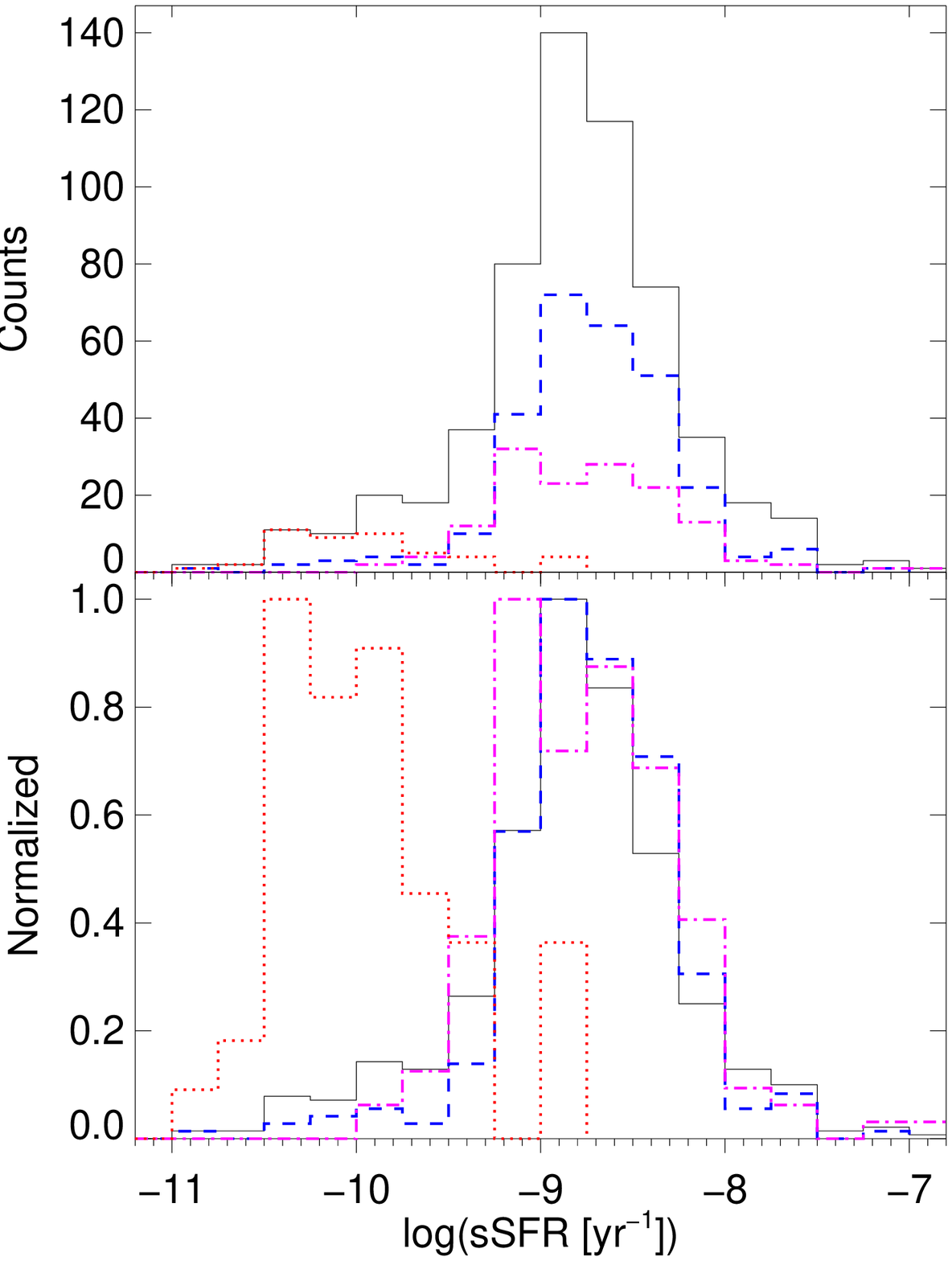}\\
    \includegraphics[width=0.33\textwidth]{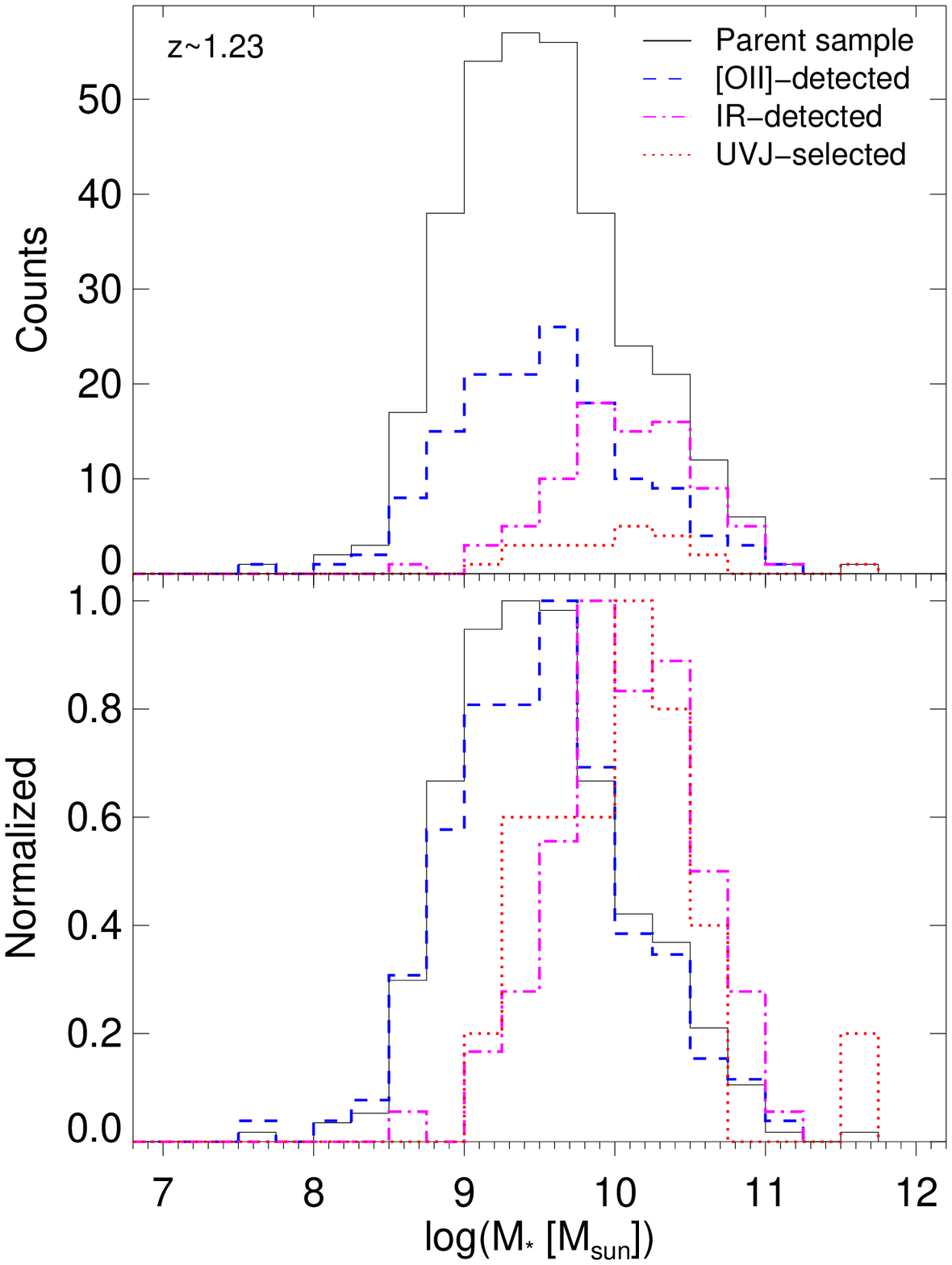}
    \includegraphics[width=0.33\textwidth]{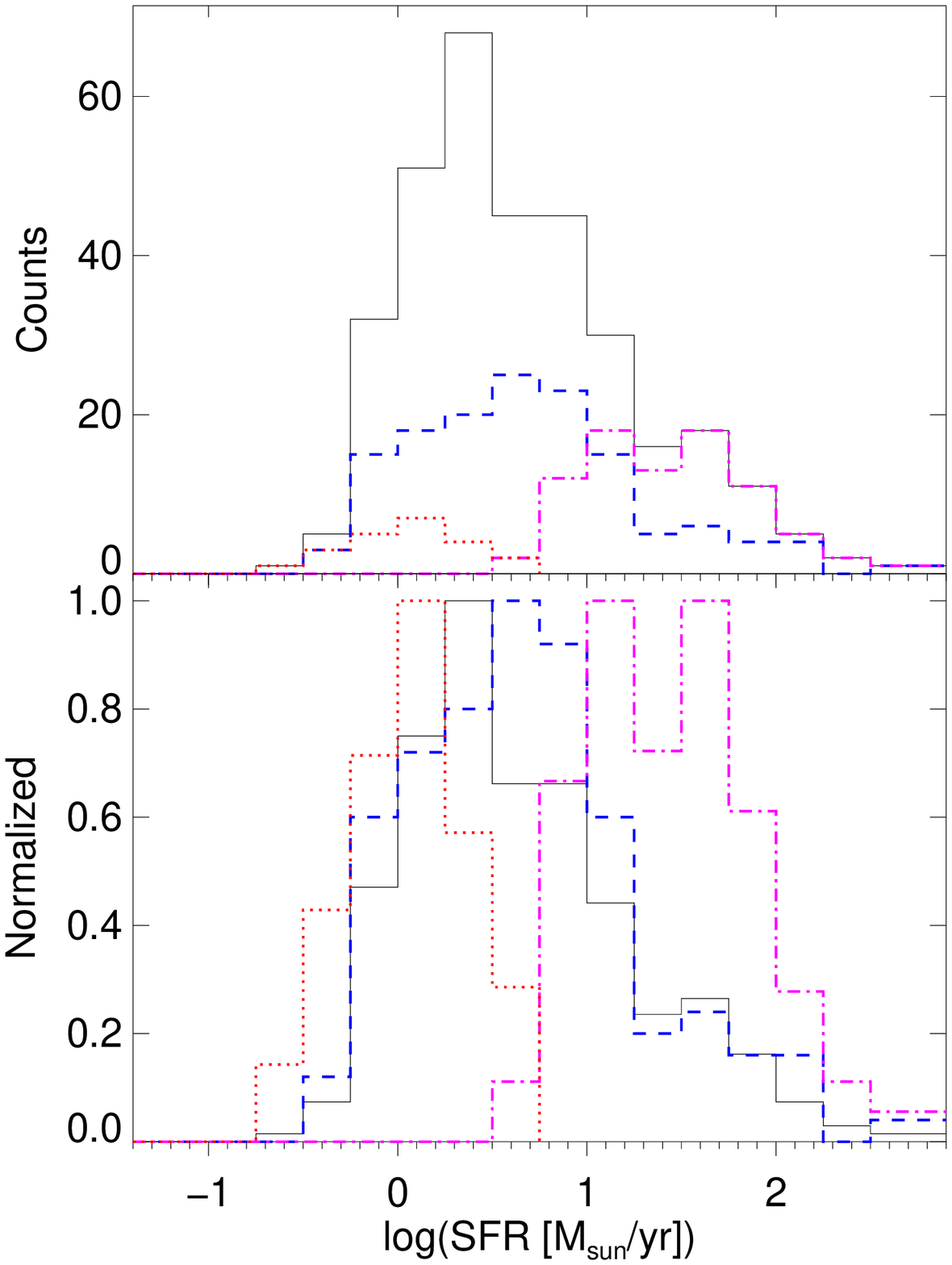}
    \includegraphics[width=0.33\textwidth]{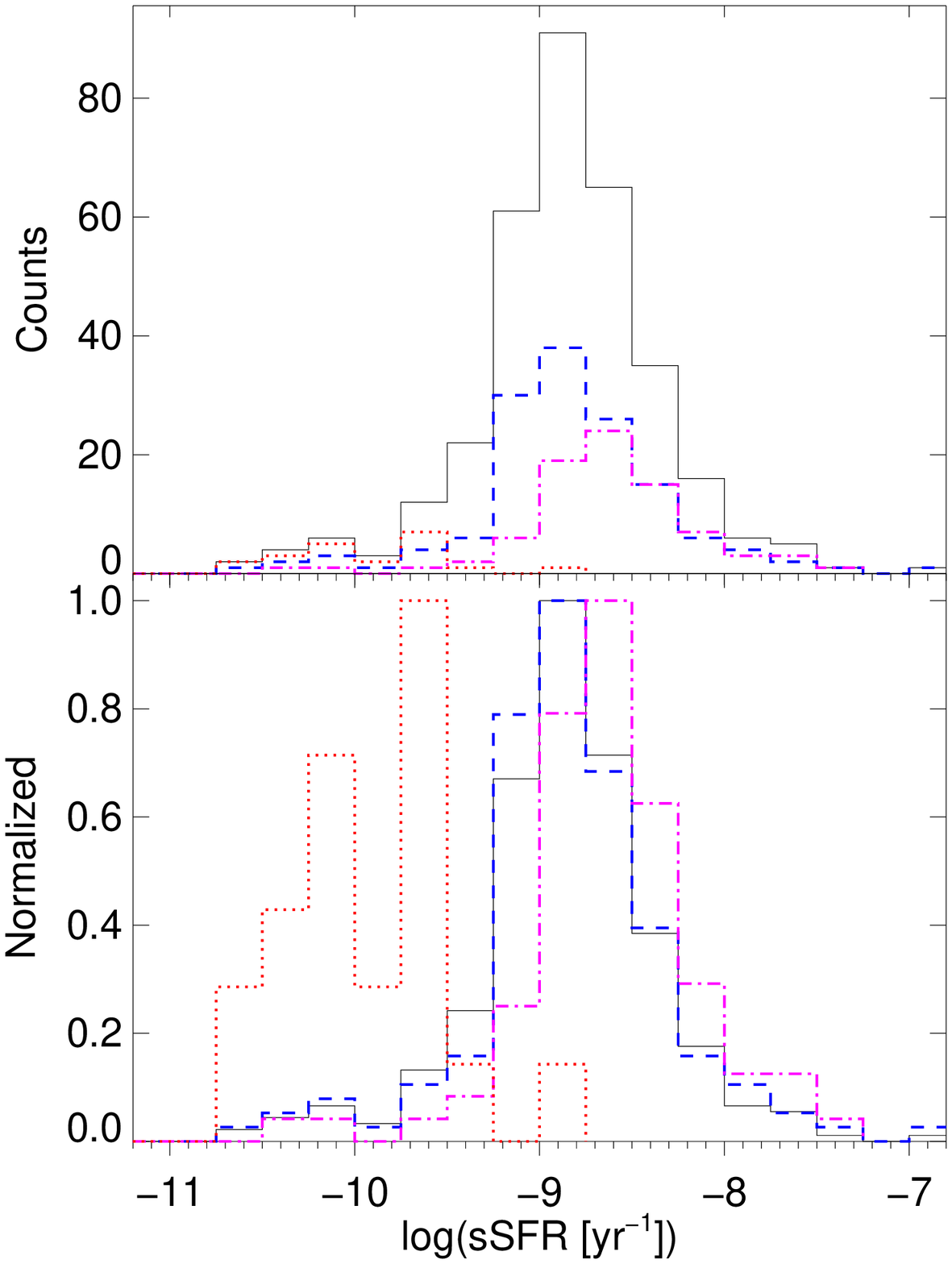}
    \figcaption{\label{fig:nf2} Distribution of M$_*$, SFR and sSFR for the z$\sim0.84$  (top panels) and z$\sim1.23$ (bottom panels) samples. Top panels in each plot represent the general number distributions, bottom panels give the histograms normalised to the peak, for each sample indicated in the legend (reported only in in the stellar mass plots).}
  \end{center}
\end{figure*} 

\section{Comparison of SFG and quiescent galaxy selections}
\label{sect:comp}
In this Section, we compare the SF properties (as derived from SED fitting) of the \OII-selected samples with the complementary samples introduced in Section~\ref{sect:complem} (nominally, the mass-deleted, the IR-detected, and the quiescent UVJ-selected samples). This allows us to investigate the effect of different sample selections on the determination of the main sequence for SFGs and to highlight the possible biases introduced by each selection.

As recently discussed in various works \citep[see, e.g.,][]{2014MNRAS.443...19R,2014ApJS..214...15S}, the determination of the SFR-M$_*$ relation using various sample definitions, related to different SFR indicators, can lead to different determination for the slope of this fundamental observed relation, with a steeper relations typically found when using UV-selected galaxies, with respect, e.g., to the IR-detected galaxy samples. This is also partially attributed to possible Malmquist bias introduced when using IR-selected galaxies, more easily detected at increasing masses.

Nonetheless, as shown in Figure~\ref{fig:nf1}, this dependence of the slope on the sample selection is not clear in our samples at redshift {\it z}\,$\sim$1. In fact, we find very similar logarithmic slopes (in all cases $\sim$1) for the SFR-M$_*$ relation of the general parent sample and the different subsamples selected on the basis of their IR and \OII\ emission (represented by the black, magenta and blue dashed lines in Figure~\ref{fig:nf1}). For completeness, we also plot the best-fit linear relation for the subsample of quiescent (UVJ-selected) galaxies, which relies well below the main sequence for SF galaxies.

The total and normalized distributions of M$_*$, SFRs and sSFRs, for each subsample are given in Figure~\ref{fig:nf2}. 
We observe that, at {\it z}\,=0.84, the \OII\ emitters have a very similar stellar mass distribution as the parent sample, maybe with a deficit at low masses (as can be seen from the normalized histograms in Figure~\ref{fig:nf2}). However, many galaxies (aproximately 50\%) below log(SFR [M$_{\rm sun}$~yr$^{-1}$])=0.8, i.e., 5-6 M$_{\rm sun}$~yr$^{-1}$  are not \OII\ emitters or are not detected by our survey. Since the expected SFR detection limits, derived from the 3$\sigma$ flux detection limit for our survey in each filter (blue dotted lines in the central panels of Figure~\ref{fig:multi_EW}), is log(SFR [M$_{\rm sun}$~yr$^{-1}$])$\sim$($-$0.2)--($-$0.7), i.e., $\sim$0.6--0.2~M$_{\rm sun}$~yr$^{-1}$, we conclude that this fraction of undetected low-SF galaxies are not due to a detection threshold, but have their \OII\ emission prevented/quenched by some physical mechanism, e.g. low excitation due to the weak SF activity. In the case of IR emitters, it is clear that they concentrate in the high-mass end of the population, and probably the detection limits, of $\sim$5~M$_{\rm sun}$~yr$^{-1}$ (shown magenta dotted in Figure~\ref{fig:nf2}), avoids detecting low SF galaxies. A similar result is seen at {\it z}\,=1.23.

\begin{figure*}[tb]
  \begin{center}
    \includegraphics[width=1.\textwidth]{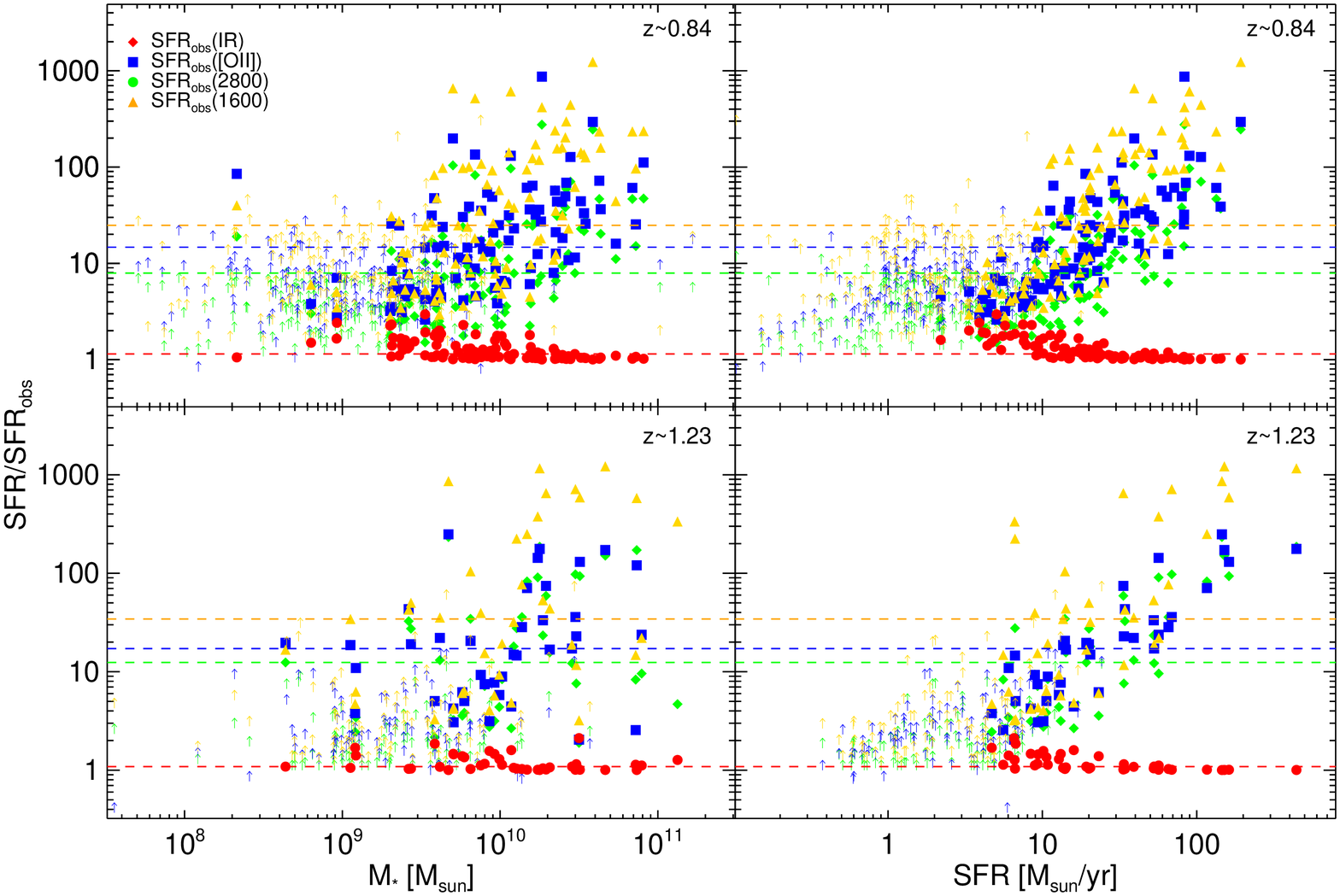}
    \figcaption{\label{fig:EXTcomp0} {\it Top panels:} ratio  between the total SFR and the observed SFR$_{\mathrm{obs}}$ for UV(1600), UV(2800), \OII\ and IR SFR indicators versus the stellar mass (left panel) and the total SFR (right panel) at redshift $\sim0.84$ (F687W17). Arrows represent lower limits for the galaxies not detected in the IR. Dashed lines represent the median values computed for IR-detected galaxies, only. {\it Bottom panels}: the same for galaxies at redshift $\sim$1.23 (F823W17).}
  \end{center}
\end{figure*}

The distributions of sSFR for the different samples are given in the right panels of Figure~\ref{fig:nf2}. Interestingly, there is good agreement between the global sSFR distribution of the parent, \OII-selected, and IR-selected samples of galaxies ). This similarity of the sSFR distributions, is mirrored by the presence of very similar main sequences for the various samples (see Figure~\ref{fig:nf1}) with a quite constant slope. Other recent studies have argued the presence of a knee in SFR-M$_*$ relation, but we do not see it in our samples \cite[see, e.g.,][]{2014ApJ...795..104W,2015ApJ...801...80L,2015arXiv150403005K}. As expected, the UVJ-selected sample of quiescent galaxies exhibits a distribution of sSFR with very small values with respect to the other sample selections.

Observing the normalised distributions, shown in the lower panels of each row of plots in Figure~\ref{fig:nf2}, we can conclude that there is not (or only mild) mass-selection bias in the case of \OII\ emitters, while the IR-detected and UVJ-selected samples are systematically shifted toward higher masses. The peak in the SFR distribution is only slightly enhanced for the \OII\ sample, with respect to the parent sample, whereas the peaks for the IR-detected and UVJ-selected samples are clearly shifted towards higher and lower SFRs, respectively. Similar conclusions apply to both redshift ranges.

As already discussed in Section~\ref{sect:complem}, the assumed sSFR cut at log(sSFR [yr$^{-1}$])=$-$9.7 (shown as dark-grey dashed lines in Figure~\ref{fig:nf1}) lead to a good overlap between the UVJ selection criteria and a pure sSFR selection for the lower redshift sample, while a slightly higher sSFR cut (log(sSFR [yr$^{-1}$])=$-$9.5, shown as dark-grey dotted line in the right panel of Figure~\ref{fig:nf1}) would better reconcile the two definitions (i.e. UVJ-selected and sSFR-based) of {\it quiescent galaxies} for the higher redshift sample. This difference might be related to a small evolution of the SF properties between these two redshift intervals (corresponding to $\sim$1.5~Gyr in time). 

Indeed we remark that a relatively large fraction of galaxies selected as quiescent using these two alternative definitions are detected as \OII\ or IR emitters. In fact we find $\sim$20\% and $\sim$25\% of the lower redshift galaxy sample to have  a \OII\ or IR detection, for the UVJ- and sSFR-selected samples, respectively. These fractions increase to  $\sim$30\% and $\sim$40\%, for the higher redshift sample. These results show, on one side, that the UVJ-selection seems to be more effective in the selection of {\it bona fide} quiescent galaxies with respect to a simple cut in the sSFR, on the other that  moving to higher redshifts the contamination from SF galaxies start to be very relevant. This result is in agreement with recent results on the study of quiescent galaxies selected in the UVJ plane \citep[see e.g.][]{2015ApJ...799..206B}.

\section{Dust attenuation properties}
\label{sect:dust}
Previous studies have found that the dust attenuation is a strong function of stellar mass (e.g. \citealt{2006ApJ...653.1004R,2010ApJ...712.1070R,2009ApJ...698L.116P,2011ApJ...742...96W,2012ApJ...754L..29W}), where the most massive galaxies are more highly obscured. 

We parametrize the attenuation as the ratio of the total SFR (see Section~\ref{sect:oii-sfr} for the detailed definition) over the observed SFR$_{\mathrm{obs}}$ for each SF indicator (UV1600, UV2800, \OII\ and IR respectively). We find a similar trend using the SHARDS samples of \OII-selected galaxies in Figure~\ref{fig:EXTcomp0}: more massive galaxies have larger SFR/SFR$_{\mathrm{obs}}$ ratios. We depict as solid symbols IR-detected galaxies for which the determination of the total SFR is more reliable, while arrows represent galaxies selected on the basis of their \OII\ emission but lacking IR-detection which may lead to an underestimation of the total SFR. Indeed, by selecting \OII\ ELGs we are possibly biasing the estimated average attenuation values toward lower values, since galaxies suffering strong dust attenuation should have their \OII\ emission reduced or non-detectable.

The IR alone is a good estimator of the total SFR for most galaxies, but there is fraction of them which have non-negligible amounts of dust-free star formation, and that seems to be more significant for lower masses. There is approximately a 20\% (25\%) of galaxies detected in the IR for which the contribution to the total SFR from unobscured star formation is larger than 30\% at redshift $\sim$0.84 ($\sim$1.23), most of them being less massive than 10$^{10}$~M$_{\rm sun}$, and with total SFR lower than 20~M$_{\rm sun}$~yr$^{-1}$. 

An analogous behaviour for the attenuation is found with respect to the total SFR (right panels in Figure~\ref{fig:EXTcomp0}), where the trend appears even more evident and the relation tighter exhibiting a smaller scatter. Galaxies sustaining higher SFRs also exhibit larger dust attenuation \citep[coherently with earlier results for SFGs, e.g.][]{2012MNRAS.426..330D,2014MNRAS.441....2D}. This characteristic may be related to the fact that more actively SFGs possibly hold higher fractions of dust sustaining this high rate of SF or could be ascribed to a higher efficiency in heating the dust in these galaxies. The median value of the ratio SFR/SFR$_{\mathrm{obs}}$, where SFR$_{\mathrm{obs}}$  is estimated from the UV(1600), UV(2800), \OII\  or IR observed (i.e. not extinction-corrected) SFR respectively, is found to smoothly increase as the wavelength of the SFR indicator decreases. 

Interestingly, the trend of the M$_*$-SFR/SFR$_{\mathrm{obs}}$ relation in the two redshift bins considered does not seem to vary significantly suggesting no (or weak) evolution of the dust attenuation properties between these two epochs (separated by $\sim$1.5~Gyr in time).
Indeed, the attenuation of the line and continuum photons as represented by the \OII\ and UV(2800) SF indicators are quite similar, again suggesting that \OII\ can be considered a reliable SF indicator. 

The results presented in Figure~\ref{fig:EXTcomp0} also suggest that galaxies of different masses are dominated by different mechanism of SF. The SFR is mostly dominated by the IR emission, when this is detectable, especially for more massive galaxies. This is indicated also by the tightening of the SFR/SFR$_{IR}$ around 1 for high masses. Massive galaxies may present violent and explosive processes of SF (traced by high dust fractions), while less massive galaxies would undergo more relaxed episodes of SF. We can interpret these results as indicative of highly unstable SF processes dominating in the massive galaxies, e.g. driven by galaxy interactions/mergers or violent relaxation in highly unstable (clumpy) disks. Less massive galaxies would instead be characterized by a more diffuse SF. The investigation of the relation between morphology/environment and SF properties is beyond the scope of this paper and is deferred to a dedicated work.


\section{Summary and conclusions}
\label{sect:concl}

We have presented a detailed analysis of a complete sample of SFGs at {\it z}\,$\sim$0.84 and {\it z}\,$\sim$1.23. The sample is a comprehensive compilation of sources selected with 3 different SFR tracers: the \OII\ emission-line, the UV emission, and the MIR/FIR emission. Concerning \OII\ emitters, we have selected them down to very faint magnitudes ($\sim$26.5~mag) benefitting from the ultra-deep spectro-photometric data from the Survey for High-z Absorption Red and Dead Sources \citep[SHARDS, see][]{2013ApJ...762...46P}. The sample of \OII\ emitters has been compared with the general population of stellar mass-selected galaxies (including from quiescent to starburst galaxies) at the mentioned redshifts. The SF activity and extinction properties of the whole population of galaxies at {\it z}\,$\sim$0.84 and {\it z}\,$\sim$1.23 have been characterized with specific SFR estimations, as well as UVJ colors and dust emission measurements based on {\it Spitzer} and {\it Herschel} data. 

The SHARDS data, combined with a customized and refined method for selection of ELGs using its medium-band ($\sim150$~\AA) filters and for the determining the continuum on the basis of SED-fitting, are suitable for ELG studies and have allowed us to obtain the results summarised below:

\begin{enumerate}
	\item we have developed a technique that optimally exploits the medium-band characteristics of the ultra-deep SHARDS survey to perform for the first time a systematic and robust detection of emission lines in low-, intermediate-, and high-redshift galaxies, inspired by the traditional procedures based on narrow-band imaging;
  	\item we have demonstrated that the depth and image quality of our survey allow us to recover virtually all ELGs which have been already confirmed by the deepest spectroscopic surveys carried out in the GOODS-N field. We have shown that we are able to extend these spectroscopic studies of ELGs, typically limited to R$\sim$24-25~mag, to fainter magnitudes ($R$$\sim$26-27~mag) and detect strong [OII] emitters with faint continuum emission (i.e., large EWs), as well as Ly-$\alpha$ emitters up to {\it z}\,$\sim$6 \citep{2014MNRAS.444L..68R}; 
	\item by combining all the SHARDS filters data and fitting them to stellar population synthesis models, we have shown that we can robustly measure equivalent widths and fluxes for these emission lines down to \EW$\sim$15-20\,\AA\ and F(\OII)$\sim$2-3$\times$10$^{-18}$\,erg\,s$^{-1}$cm$^{-2}$ respectively. These values are very similar to what can be done with deep spectroscopy. But, spectroscopic surveys typically pre-select their targets based on broad-band magnitudes (RI$<$24--25~mag) and have limited completeness because they can only put a limited number of slits in every object, unless huge amount of observing time is granted. By using our medium-band selection technique, we get all emitters without any strong selection effect and down to fainter broad-band magnitudes.
	\item we have focused this paper on two redshift ranges. One based on the selection of galaxies with emission lines lying within the F687W17 filter (central wavelength: 687~nm), and another sample selected with the F823W17 filter. For \OII\ emitters, these wavelengths correspond to {\it z}\,$\sim$0.84 and {\it z}\,$\sim$1.23. The use of the SHARDS-based photometric redshifts allowed to extend (a factor $\sim3$ in numbers) the samples selected based only on spectroscopic redshift confirmation available from literature. We can robustly measure \OII\ fluxes and EWs even for galaxies that typically do not have sufficiently high-quality spectra for these kind of measurements (including galaxies usually exhibiting \OII\  emission but lacking continuum detection); 
	\item the full SHARDS dataset, spanning the optical range from 500 to 950~nm, has allowed a significant improvement (about one order of magnitude, reaching $\sigma_{zphot}\lesssim0.5\%$ at redshift {\it z}\,$\sim$1) on the photometric redshift accuracy \citep[see][and G.~Barro et al. 2015, in preparation]{2013ApJ...762...46P,2014MNRAS.444..906F} with respect to the use of previously available broad-band data. The accuracy in the photo-z determination is mirrored in the high success rate ($\gtrsim$90--95\%) and low contamination levels ($\lesssim$5--10\%) of the selected samples of \OII\ emitters; 	
	\item we have analysed the \EW\ distribution for the two samples of \OII\ emitters, finding robust evidences on the evolution of the average \EW\ with cosmic time at least up to redshift $\sim$1, followed by a possible flattening, in good agreement with what is found studying the SFRD evolution of the Universe. Thus, globally speaking, the \OII\ emission is a good tracer of the SFR evolution, despite the complicated dependence on intrinsic physical parameters such as metallicity and temperature, among others; 
	\item we have found that galaxies with low-\EW\ (i.e. below the median \EW) have on average higher masses with respect to galaxies with high-\EW\ (by a factor $\sim$4 and $\sim$2 for the samples at {\it z}\,$\sim$0.84 and {\it z}\,$\sim$1.23, respectively). We do not see a similar trend with respect to the age of the stellar populations. Instead, we observe that galaxies with masses below the median mass value have on average younger ages, by a factor of $\sim$2 ($\sim$4), with respect to the more massive galaxies for the {\it z}\,$\sim$0.84 ({\it z}\,$\sim$1.23) sample. As might be expected from the downsizing scenario, massive galaxies seem to have formed the bulk of their stellar mass at an earlier epoch;
	\item we derive the SFR(\OII) and compare the SFR-M$_*$ relation for our samples with previous results from literature. We find a steeper slope ($\alpha\sim$0.9) of the relation for both redshift intervals with respect to \cite{2012ApJ...754L..29W}, but this is compatible with their results when selecting blue galaxies (or in the case where we adopt ad mass-dependent extinction calibration). Additionally, our data suggest a trend of the slope with redshift, with higher redshift galaxies showing shallower slopes (despite the various caveats related to possible observational biases). Indeed the slope of the SFR-M$_*$ relation for our \OII\ emitters is in very good agreement with the main sequence of {\it z}\,$\sim$1 SF galaxies defined in \cite{2007A&A...468...33E}. Also the scatter in the relation ($\sim$0.35--0.38~dex) is in agreement with previous determinations;
	\item we have compared the distributions of physical properties and the SFR-M$_*$ relation for the \OII- and IR-detected samples with those of the general population of galaxies at the same redshift (parent sample), finding no evident bias introduced on the SFR-M$_*$ relation by any of the selections of SFGs. Quiescent galaxies (identified on the basis of the UVJ diagram or a sSFR cut), populate a separated locus in the SFR-M$_*$ diagram and follow an independent relation, although there is a large scatter in the data points and some outliers with higher SFRs;
	\item we have discussed the possible biases affecting the different samples, comparing the total and normalised distributions for the M$_*$, SFRs and sSFRs, for the subsamples of \OII-selected, IR-detected and UVJ-selected quiescent galaxies, in the two selected redshift bins. We find only a very mild trend in selecting more massive galaxies based on the detection of the \OII\ emission line. A stronger bias is found relatively to the M$_*$ and SFR distributions of IR-detected galaxies, and (as expected) for the quiescent sample. We ascribe the larger bias for the IR-detected samples of galaxies to their larger mass- and SFR-detection limits;
	\item we observe very similar sSFR distributions for the parent sample and the \OII- and IR-detected samples of SF galaxies. This behaviour is in relation to the very similar SFR-M$_*$ relations found for the various samples. We dot not observe significant variations in  the slope of the main sequence of SF galaxies at high masses (M$_*>10^{10}$M$_{\rm sun}$), as claimed by other recent works;
	\item we observe that the sample of quiescent galaxies, independently from the definition based on UVJ diagram or a sSFR cut, can be contaminated by both \OII\ and IR-detected galaxies. We find that $\sim$20\% and $\sim$25\% of the lower redshift galaxy sample present \OII\ or IR detection, for the UVJ- and sSFR-selected samples respectively. These fractions increase to  $\sim$30\% and $\sim$40\%, for the higher redshift sample. We deduce that the UVJ-selection seems to be more effective in the selection of {\it bona fide} quiescent galaxies with respect to a simple cut in the sSFR, on the other hand we note that  moving to higher redshifts the contamination from SF galaxies start to be very relevant and that more careful (redshift dependent) selection criteria should be adopted;
	\item we observe that, the IR emission alone is a good estimator of the total SFR for most galaxies, but there is fraction of them which have non-negligible amounts of dust-free star formation, and that seems to be more significant for lower masses. Approximately 20\% (25\%) of IR-detected galaxies present a contribution to the total SFR from unobscured star formation larger than 30\% at redshift $\sim$0.84 ($\sim$1.23), most of them being less massive than 10$^{10}$~M$_{\rm sun}$, and with total SFR lower than 20~M$_{\rm sun}$~yr$^{-1}$. 
	\item we have finally found that the dust attenuation (as indicated by the ratio of the total to the observed SFR for each SFR indicator) strongly correlates with the stellar mass and total SFR of the galaxies. This result might be related to a higher dust content for more massive galaxies or to an higher dust heating efficiency in these massive objects. The dust attenuation correlates linearly with stellar mass, and the trend of the stellar mass-dust attenuation relation does not show significant differences between the two redshift ranges, suggesting no (or slow) evolution of dust attenuation properties between these two epochs ($\sim$1.5\,Gyr apart). These results also suggest that galaxies of different masses are dominated by different mechanism of SF, with massive galaxies exhibiting more violent and explosive processes of SF (traced by the higher dust fractions), while less massive galaxies would undergo more moderate episodes of SF;
\end{enumerate}

We have shown the huge potential of SHARDS ultra-deep data in studying galaxy evolution using ELGs.
We plan to extend the results presented in this work using the whole SHARDS dataset (including 25 filters and several detected lines) to constrain the evolution of the physical properties of SFGs in the redshift interval $0.3<\,{\it z}\,<2$, fully covering the epoch where strong evolutionary processes shaped the galaxy populations toward the present-day Universe.

\section*{Acknowledgments}

The authors would like to thank the Referee for her/his valuable comments and helpful suggestions concerning the presentation of this paper, which helped to improve the manuscript.
The work of A.C. is supported by the STARFORM Sinergia Project funded by the Swiss National Science Foundation, and also benefited from a MERAC Funding and Travel Award. 
A.C. acknowledges the kind hospitality of the Departamento de Astrof\'{i}sica y Ciencias de la Atm\'{o}sfera, Universidad Complutense de Madrid, and of the Departament d?Astronomia i Astrofisica, Universitat de Valencia. 
A.C. and E.R. are grateful to M.~Cava for many useful discussions and suggestions. 
E.R. acknowledges the kind hospitality of the  Observatoire de Gen\`{e}ve, Université de Gen\`{e}ve. 
P.G.P.-G. acknowledges support from the Spanish Programa Nacional de Astronom\'{\i}a y Astrof\'{\i}sica under grants AYA2012-31277. 
A.V.G. acknowledges support from the ERC via an Advanced Grant under grant agreement no. 321323-NEOGAL. 
N.C. acknowledges support from the Spanish Ministry of Economy and Competitiveness under grant AYA2013-
46724-P.
A.A.H. and A.H.C. acknowledge  financial support from the Spanish Ministry of Economy and Competitiveness through grant AYA2012-31447, which is partly funded by the FEDER program. 
A.J.C. is a Ram\'on y Cajal Fellow of the Spanish Ministry of Science and Innovation and acknowledges financial support from the Spanish Ministry of Economy and Competitiveness through grant AYA2012-30789, partly funded by the FEDER program. 
This work has made use of the Rainbow Cosmological Surveys Database, which is operated by the Universidad Complutense de Madrid (UCM). Based on observations made with the Gran Telescopio Canarias (GTC), installed at the Spanish Observatorio del Roque de los Muchachos of the Instituto de Astrof\'{\i}sica de Canarias, in the island of La Palma.

\bibliographystyle{apj}
\bibliography{referencias}

\label{lastpage}
\end{document}